\documentclass[preprint,aps]{revtex4}
\usepackage{epsfig}
\begin{document}

\title{Electron Scattering on \boldmath{${}^3$}He 
       - a Playground to Test Nuclear Dynamics
}
\altaffiliation[]{Presented at $6^{th}$ Workshop on ``e-m induced Two-Hadron Emission'',
Pavia, 2003}

\author{W.~Gl\"ockle}
\affiliation{Institut f\"ur Theoretische Physik II,
         Ruhr Universit\"at Bochum, D-44780 Bochum, Germany}

\author{J.~Golak, R.~Skibi\'nski, H.~Wita{\l}a}
\affiliation{M. Smoluchowski Institute of Physics, Jagiellonian University,
                    PL-30059 Krak\'ow, Poland}
\author{H.~Kamada}
\affiliation{Department of Physics, Faculty of Engineering,
  Kyushu Institute of Technology,
  1-1 Sensuicho, Tobata, Kitakyushu 804-8550, Japan}
\author{A.~Nogga}
\affiliation{Institute for Nuclear Theory, University of Washington,
Seattle, WA 98195-1550, USA}
 
\begin{abstract}
 The big spectrum of electron induced processes on $^3$He is illustrated
 by several examples based on Faddeev calculations with modern nucleon-nucleon
 and three-nucleon forces as well as exchange currents. The kinematical
 region is restricted to 
a mostly nonrelativistic one where the three-nucleon c.m. energy is below
 the pion production 
 threshold and the three-momentum  of the virtual photon is sufficiently
 below the nucleon mass. Comparisons with available data are shown and 
cases of  agreement and disagreement are found. It is argued that new and
 precise data are needed to systematically check the present day
 dynamical ingredients.
\end{abstract}
\pacs{21.45+v,21.10-k,25.10+s,25.20-x}
\maketitle

\section{Introduction}

If one chooses such energy and momenta of the virtual photon that the
three-nucleon (3N) c.m. energy in the final state is below the pion  
 production threshold and
the total 3N momentum remains sufficiently well below the nucleon mass, 
a nonrelativistic approach is well justified. In that
kinematical region nucleon-nucleon (NN) forces 
are well tuned to the data.
Also 3N
forces are adjusted to the binding energy of the $^3$H nucleus. In
principle the interplay of those two types of  dynamical ingredients can 
and should
be tested via the rich set of
3N scattering observables. A successful test would guarantee a correct
description of the final state interaction for electron induced inelastic
processes on $^3$He. 
It would also  provide  
confidence  that the $^3$He bound state
wave function is reliable. 
Those tests are still going on using nucleon-deuteron (Nd) 
elastic scattering and breakup reactions. Though overall one can
say already  now  \cite{pure3N} that the great bulk of the 
existing 3N scattering data is quite
well described by the present day force models, the 
situation,  however, is by far not ideal because the 3N force properties 
are still rather unsettled now~\cite{3NFproblems}.
If such  an ideal situation of a successful
description of 3N scattering observables was reached, the only new
dynamical ingredients in electron scattering on $^3$He
would be the electromagnetic nucleonic current operators. 
Nowadays they still
pose a serious challenge and a generally sound parametrization has not
yet been achieved. In such a situation combined efforts, in the pure 
3N sector and in electron induced processes, appear advisable to forward 
and to lay a solid ground of data. 

\begin{figure}[!ht]
  \begin{center}
\epsfig{file=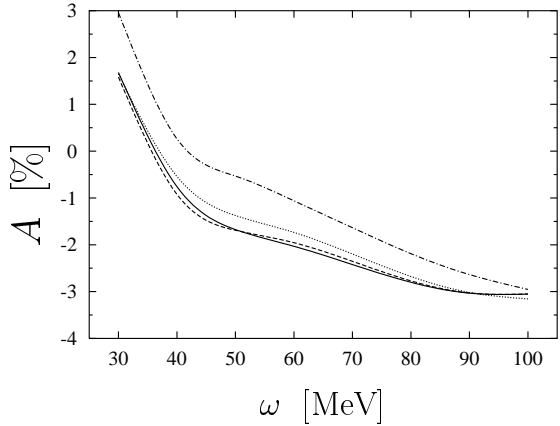,height=6cm}
    \caption{\label{fig1}
            The asymmetry $A$ 
            against the energy transfer $\omega$
            for $q^2$= 0.1 ${\rm (GeV/c)}^2$. The curves describe results 
            based on the approximate treatment of FSI (see text) 
(dash-dotted line),
            the results under full inclusion of FSI but without MEC 
(dotted line),
            the full calculations including MEC (dashed line) and
            the full calculations which incorporate both $\pi$- and 
$\rho$-like 
            MEC~\protect\cite{MEC} and 3N force 
            (here Urbana~IX~\cite{UrbanaIX}) effects (solid line).
            AV18~\protect\cite{AV18} is used as the NN potential.
      }
    \end{center}
\end{figure}

As an introductory illustrative example for electron induced processes 
on $^3$He we would like to use the asymmetry $A$ 
in inclusive scattering of a polarized 
electron beam on a polarized $^3$He target. 
We show that its description requires the full
list of dynamical ingredients. In Fig.~\ref{fig1} $A$ is displayed
as a function of the photon energy $\omega$. We see the result of an often used
approximation (dashed-dotted line) in which the photon is absorbed by just 
one nucleon and the two spectator nucleons interact in first order 
in the NN $t$-operator. (This corresponds to the standard approximation 
leading to the widely used concept
of the spectral function). Apparently for that observable this
is by far not sufficient and the final state interaction  among all three
nucleons is essential. It shifts the theory significantly downwards. Allowing 
in addition that the photon is absorbed by two-nucleon currents causes an
additional shift and even effects of 3N forces are visible. This example
clearly illustrates that there are observables for electron 
scattering on $^3$He,
which reflect the various dynamical ingredients in a sensitive manner. As
we shall see below this observable allows to extract the magnetic form
factor of the neutron, the extracted value of which would be very 
unreliable if simple minded approximations in the analysis were used.

In Sect.~\ref{sec:2} we briefly review our theoretical framework.
Then in Sect.~\ref{sec:3} we display various observables
for electron induced processes on $^3$He which show both, agreement and
disagreement with theory. The need for additional data is pointed out to
challenge the present day theory in a more specific and systematic manner
than, we think, it has been done up to now. We end with an outlook in Sect.~\ref{sec:4}.

\section{Theoretical Framework}
\label{sec:2}

The central quantities in the description of electron induced processes
are the nuclear matrix elements:
\begin{equation}
N^\mu \equiv
\langle \Psi_f^{(-)} \mid j^\mu ( \vec Q) \mid 
\Psi^{\theta^* \phi^*}_{{}^3{\rm He}}
\rangle .
\label{eq1}
\end{equation}
They are composed of the polarized $^3$He target state,
the components $j_\mu(\vec Q)$ 
of the current operator and the final 3N scattering state 
$ \langle \Psi_f^{(-)} \mid $
with asymptotic momenta and spin quantum numbers $f$. 
For proton-deuteron (pd) breakup these are the proton
and deuteron momenta and their spin magnetic quantum numbers, and for 
the full breakup the three final nucleon momenta and again their 
spin magnetic quantum numbers.
In the latter case one has to add the isospin labels.
The initial $^3$He spin direction is determined by the angle $\theta^*$ with respect 
to the photon momentum $\vec Q$ and by the azimuthal angle $\phi^*$
in relation to the scattering plane formed by the initial and final electron momenta.

The $^3$He state is nowadays for instance a straight solution 
\cite{Nogga} of the Faddeev equation
\begin{equation}
\psi = G_0 t P \psi + G_0 ( 1 + t G_0 ) V_4^{(1)} ( 1 + P ) \psi ,
\label{eq2}
\end{equation}
where
the Faddeev component $\psi$  determines the full state via
\begin{equation}
\Psi_{{}^3{\rm He}} = ( 1 + P ) \psi .
\label{eq3}
\end{equation}
The ingredients are the free 3N propagator $G_0$, 
the NN t-operator generated via
the Lippmann-Schwinger equation from any modern NN interaction 
and a suitably chosen permutation operator $P$~\cite{gloecklebook}. 
Further $V_4^{(1)}$ is one of the  three parts
of a 3N force into which any 3NF can  be decomposed. We assume here 
that the $t$-operator acts in the pair 23
and that $V_4^{(1)}$ is the part of the three-nucleon force 
which is symmetrical under exchange of particles 2 and 3.

It is advisable not to
evaluate the scattering states separately but to redirect the action of
the M\"oller wave operator towards the current and the target state. 
This leads \cite{photon1} to the following form of the nuclear matrix element
\begin{equation}
N^\mu = \langle  \psi_f \mid U^\mu \rangle ,
\label{eq4}
\end{equation}
where 
the auxiliary state $ \mid U^\mu \rangle $ 
obeys the Faddeev-like integral equation
\begin{eqnarray}
\mid U^\mu \rangle 
& = & (1 + P ) j^\mu ( \vec Q ) \mid  \Psi_{{}^3{\rm He}} \rangle \cr
& + & P t G_0 \mid U^\mu \rangle + ( 1 + P ) V_4^{(1)} G_0 ( t G_0 + 1 ) \mid U^\mu \rangle .
\label{eq5}
\end{eqnarray}
$ \vert \psi_f \rangle$ is a known channel-dependent state, 
which for the pd case is just the deuteron state
together with a plane wave for the third particle, 
$\vert \psi_f^{pd} \rangle = \vert \phi_d {\vec q} \rangle$.
For the complete three-body breakup it is given as  
$\vert \psi_f^{3N} \rangle = (1 + G_0 t) \vert \phi_0 \rangle$,
with $\vert \phi_0 \rangle $ being plane waves, antisymmetrized 
in the two-body subsystem, where $t$ acts.

\begin{figure}[htb]
  \begin{center}
\epsfig{file=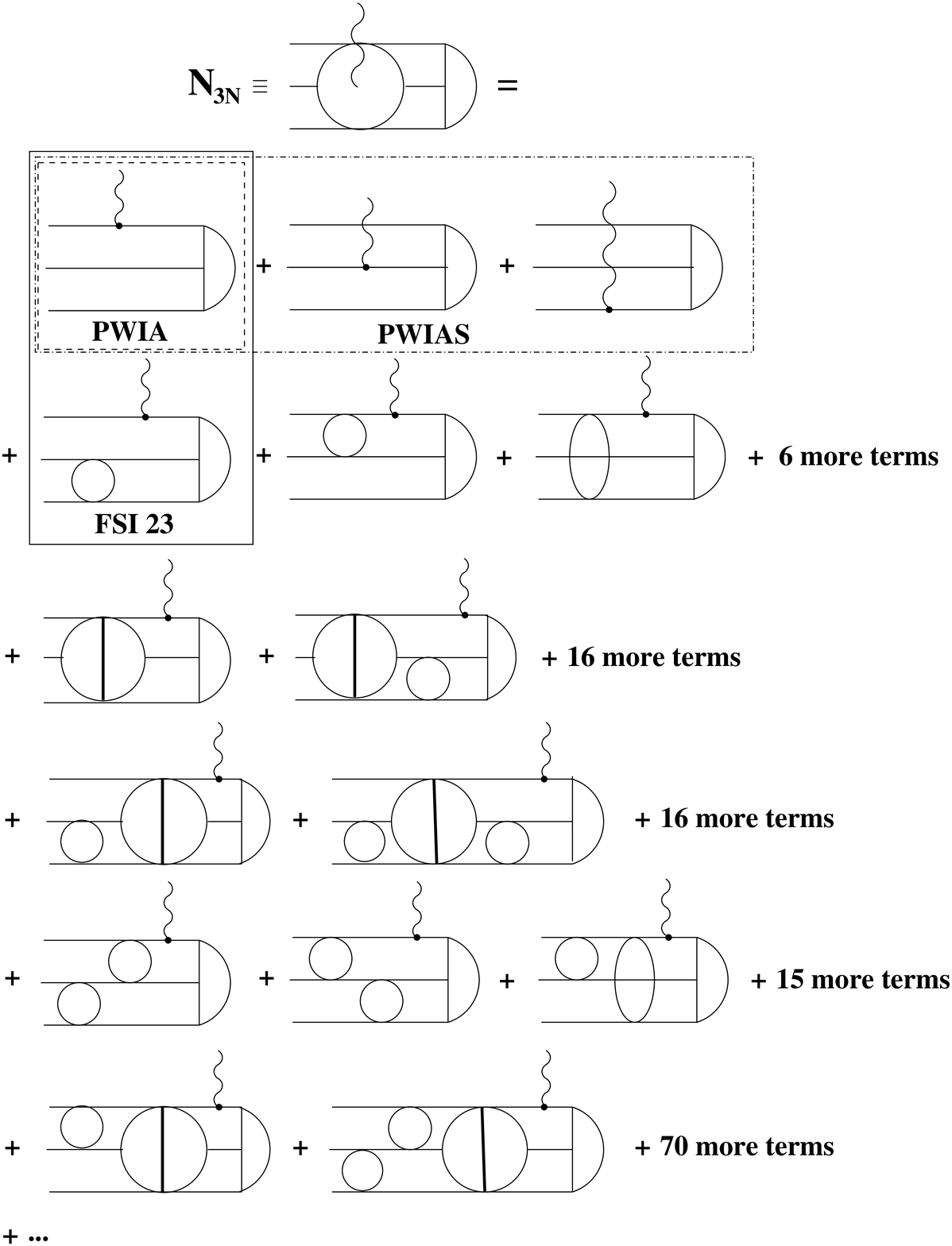,height=9cm}
    \caption{\label{fig2}
Diagrammatic representation of the nuclear matrix element 
for the three-body electrodisintegration of $^3$He.
The open circles and ovals represent the two-body $t$-matrices.
The big circles with a vertical line denote the action of $V_4^{(1)}$. 
Three horizontal lines between photon absorption and forces, and between 
forces describe free propagation. The half-moon symbol on the very right
stands for $^3$He.
      }
    \end{center}
\end{figure}

We see in Eq.~(\ref{eq5}) a similar type of Faddeev-like equation 
as for the bound state,
but now there are singularities in $t$ and $G_0$, which have 
to be treated appropriately.
This equation is not suitable for numerical applications (except for
forces of finite rank) because the permutation operator stands to the
very left \cite{Ubad}. The equation can be rewritten
to an appropriate form suitable for
numerical implementation \cite{photon1}. Here, in order to display the
physical content of the matrix element it is, however, quite adequate. To that aim we
iterate Eq.~(\ref{eq5}). In obvious notation it reads
\begin{equation}
\mid U^\mu \rangle =
\mid U_{0}^\mu \rangle + K_{NN} \mid U^\mu \rangle
+ K_{3N} \mid U^\mu \rangle
\label{eq6}
\end{equation}
and after iteration
\begin{eqnarray}
& \mid U^\mu \rangle = & \mid U_{0}^\mu \rangle +   \nonumber \\
& & ( K_{NN} +  K_{3N} ) \mid U_{0}^\mu \rangle + \nonumber \\
& & ( K_{NN} +  K_{3N} ) ( K_{NN} +  K_{3N} ) \mid U_{0}^\mu \rangle + \cdots
\label{eq7}
\end{eqnarray}
The resulting terms building  up $N^\mu$ are depicted in Fig.~\ref{fig2}
for the case of complete breakup.

In general the series shown in Fig.~\ref{fig2} converges very slowly 
or even diverges~\cite{inclusive1,more4}.
Therefore it is important to rely on
the full solution of Eq.~(\ref{eq5}), which also guarantees that  
$ \langle  \Psi_f^{(-)} \mid $  and $ \mid \Psi_{^3{\rm He}} \rangle $ are
consistent solutions to the same 3N Hamiltonian.

For the current operator we use the standard nonrelativistic 
single nucleon piece 
and two-body currents 
of the $\pi$ -
and $\rho$ - exchange type related to the AV18 NN force~\cite{AV18}
as proposed by Riska~\cite{MEC}. In case of
elastic electron scattering on $^3$He the charge form factor 
is known \cite{chargeff} to be
sensitive to an additional two-body density operator and we use the
seagull terms from \cite{charge2b}.

\begin{figure}[htb]
  \begin{center}
\epsfig{file=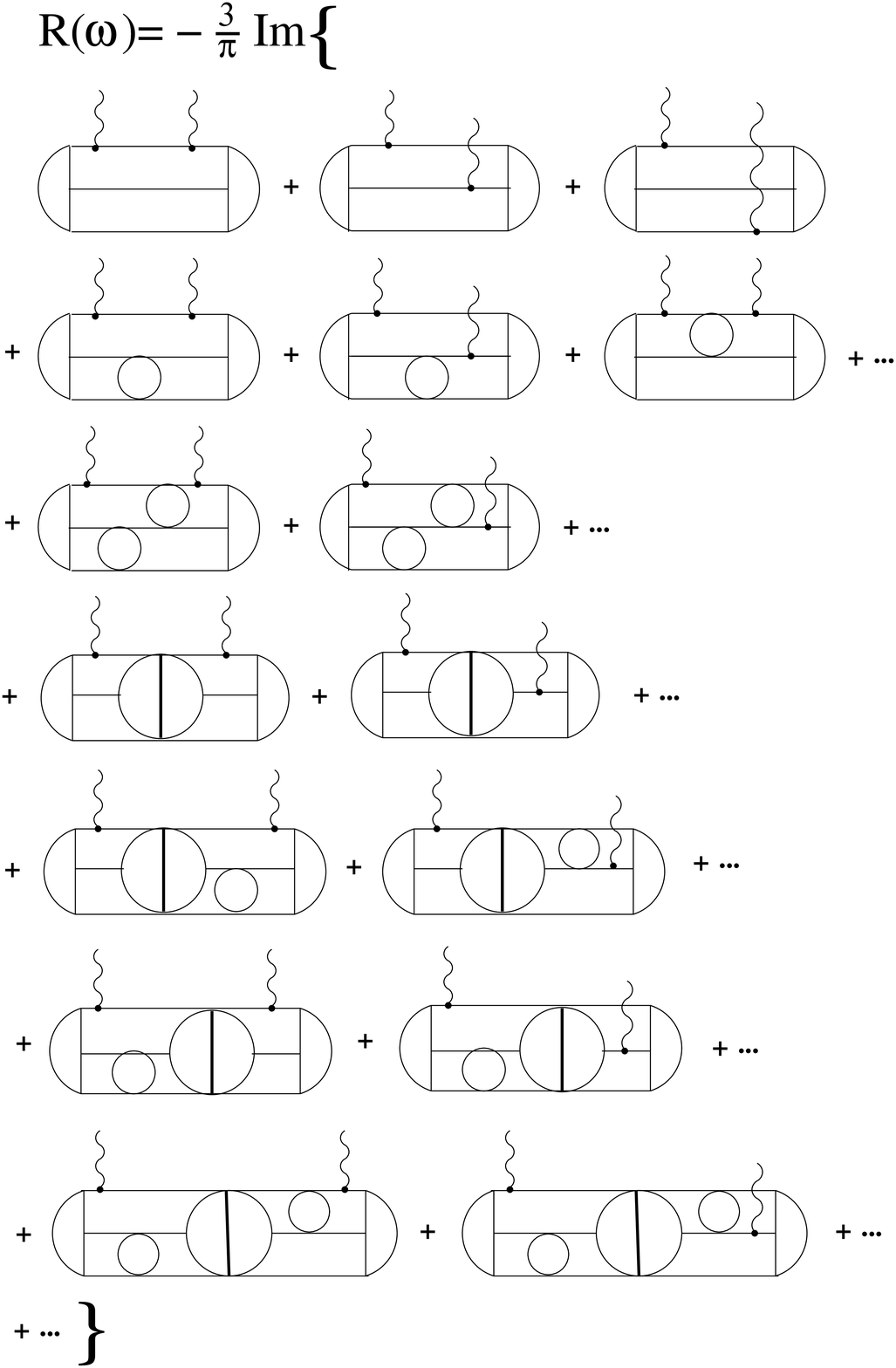,height=10cm}
    \caption{\label{fig2a}
The processes contributing to the inclusive response function $R(\omega)$ from 
Eq.~(\protect\ref{eq8}), where ${\hat O}$ is photon absorption on a single 
nucleon. Symbols as in Fig.~\protect\ref{fig2}.
      }
    \end{center}
\end{figure}

The formulation just presented is applicable for all sorts of inelastic
 electron induced processes on $^3$He. 
In case of inclusive processes one can also
 use an alternative approach which is based on the closure relation. 
The generic 
form for a response function in inclusive processes is
\begin{eqnarray}
R (\omega) =
\sum\limits_f
\left| \langle \Psi_f^{(-)} \mid \hat O \mid \Psi_i \rangle \right|^2
\delta ( \omega + E_i - E_f ) \nonumber \\
= \sum\limits_f
\langle \Psi_i \mid {\hat O}^\dagger \mid \Psi_f^{(-)} \rangle
\delta ( \omega + E_i - E_f )
\langle \Psi_f^{(-)} \mid \hat O \mid \Psi_i \rangle \nonumber \\
= \sum\limits_f
\langle \Psi_i \mid {\hat O}^\dagger
\delta ( \omega + E_i - H )
\mid \Psi_f^{(-)} \rangle \langle \Psi_f^{(-)} \mid \hat O \mid \Psi_i \rangle
\nonumber \\
= \langle \Psi_i \mid {\hat O}^\dagger
\delta ( \omega + E_i - H )  \hat O \mid \Psi_i \rangle
\nonumber \\
= - \frac1{\pi} \Im \left(
\langle \Psi_i \mid {\hat O}^\dagger
\frac1{ E + \epsilon - H }  \hat O \mid \Psi_i \rangle \right) ,
\label{eq8}
\end{eqnarray}
where $E$ is now the internal c.m. 3N energy.
This suggests to define an auxiliary state
\begin{equation}
\mid \Psi_{\hat O} \rangle \equiv \frac1{ E + \epsilon - H } \hat O
\mid \Psi_i \rangle .
\label{eq9}
\end{equation}
That auxiliary state $\mid  \Psi_{\hat O} \rangle $ 
fulfills an inhomogeneous Schr\"odinger equation 
\begin{equation}
( H - E ) \mid \Psi_{\hat O} \rangle =  {\hat O} \mid \Psi_i \rangle ,
\label{eq10}
\end{equation}
which can again be solved precisely by a Faddeev-like formulation. 
It results in
\begin{equation}
\mid \Psi_{\hat O} \rangle = G_0 ( 1 + P ) U ,
\label{eq9_a}
\end{equation}
where U obeys
\begin{equation}
U = ( 1 + t G_0 ) O^{(1)} \mid \Psi_i \rangle  + t G_0 P U 
+ ( 1 + t G_0 ) V_4^{(1)} G_0 ( 1 + P ) U 
\label{eq9_b}
\end{equation}
and the operator $\hat O$ has been decomposed as 
 $\hat O = O^{(1)} + O^{(2)} + O^{(3)}$, which is always possible. 
 We refer to \cite{inclusive1} and to \cite{inclusive2} 
where also  the slightly more intricate case with polarized 
particles is described. Again we would like to graphically illustrate the
physical content, now for the final expression of Eq.~(\ref{eq10}). 
Iterating the Faddeev-like equation (\ref{eq9_b}) 
 for the quantity $U$ 
one obtains the series of processes depicted in Fig.~\ref{fig2a}. 

Our numerical accuracy in the observables is of
the order of one percent.
It should also be noticed that all dynamical ingredients are fixed and no
free parameters are used (except in the case that the neutron form
factors
are determined). If not otherwise stated we always use in this paper 
the AV18 NN force, 
the Urbana IX 3NF \cite{UrbanaIX} and the currents as mentioned above.
   

Usually the vector part  of $N^\mu$ is decomposed into a component 
parallel to the
photon momentum $\vec Q$ and one  transversal to it, where the parallel 
component
is
eliminated with the help of the continuity equation  in favor of the
density matrix element $N^0$. 
This leads to the standard expressions for the
cross sections \cite{formula.xs1,formula.xs2} 
in terms of response functions $R_i$, the 
 analytically known kinematical factors $v_i$, and 
the electron beam helicity  $h$:

\begin{eqnarray}
d \sigma = \sigma_{\rm Mott} \, 
\frac1{(E_{e'})^2} \, \delta^4 (P_f - P_i - Q) \,
d^3 k_{e'} \, d^3 p_1 \, d^3 p_2 \, d^3 p_3 \, \nonumber \\
\left[  v_L R_L + v_T R_T + v_{TT} R_{TT} + v_{TL} R_{TL}
   + h \left( v_{TL'} R_{TL'} + v_{T'} R_{T'} \right) \right] .
\label{eq11}
\end{eqnarray}

Thus the eightfold differential cross section 
$ d ^8 \sigma / ( d E_{e'} d \Omega_{e'} d \Omega_{1} d \Omega_{2} d E_1 ) $
 for complete breakup reads 
\begin{eqnarray}
& \frac{d ^8 \sigma}{ d E_{e'} d \Omega_{e'} d \Omega_{1} d \Omega_{2} d E_1 } = &
\nonumber \\
& \sigma_{\rm Mott} \, \left[  v_L R_L + v_T R_T + v_{TT} R_{TT} + v_{TL} R_{TL} \right.  &
\nonumber \\
& \left.
   + h \left( v_{TL'} R_{TL'} + v_{T'} R_{T'} \right) \right] \, \rho_f , &
\label{eq11.5}
\end{eqnarray}
where the phase space factor $\rho_f$ in terms of final nucleon laboratory momenta 
${\vec p}_i$ and the nucleon mass $m$ is
\begin{equation}
\rho_f = 
\frac{m p_1 p_2^2}{\left| \frac{p_2}{m} 
- \frac{ {\vec p}_2 \cdot {\vec p}_3  }{m p_2} \right| } .
\label{eq12}
\end{equation}
Note that the matrix elements
which appear in $R_i$ in Eqs.~(\ref{eq11}) and~(\ref{eq11.5})
are calculated for a polarized initial $^3$He state.
If the final polarizations are not measured, we sum the response 
functions over the
magnetic quantum numbers of the three final nucleons.
In particular, 
\begin{equation}
R_L \equiv \sum\limits_{m_1,m_2,m_3} 
\left| 
N^0 ({\vec p}_1, {\vec p}_2, {\vec p}_3 ; 
m_1,m_2,m_3; \nu_1,\nu_2, \nu_3 ; \theta^*, \phi^*
) 
\right|^2 ,
\label{eq135}
\end{equation}
where $m_1, m_2, m_3$ are spin magnetic 
quantum numbers, and $ \nu_1,\nu_2, \nu_3 $ are isospin magnetic 
quantum numbers needed 
to identify the nucleons in the final state.


\section{Selected Observables}
\label{sec:3}

We will cover the various types of observables by selecting a few
examples. A previous overview of similar type but based on less 
complete dynamical ingredients
 was given before in~\cite{review}.
As emphasized in the introduction the kinematics is restricted  such that a
nonrelativistic treatment appears justified.  

\subsection{Elastic Electron Scattering}

\begin{figure}[htb]
  \begin{center}
 \epsfig{file=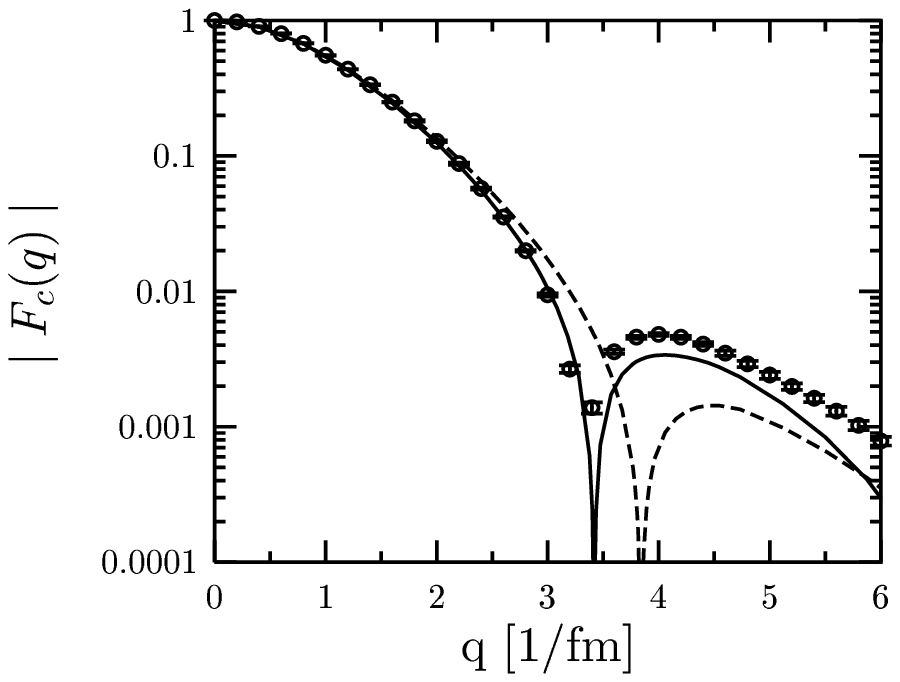,height=4.5cm}
 \epsfig{file=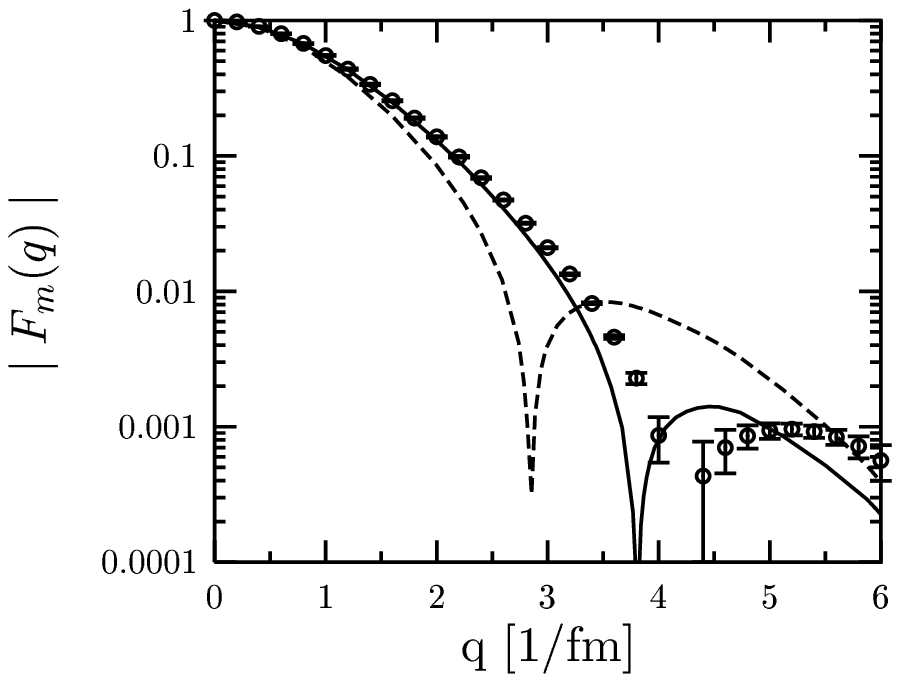,height=4.5cm}
    \caption{\label{fig4}
The elastic charge ($\vert F_c(q)\vert$) and 
magnetic ($\vert F_m(q)\vert$) form factors of $^3$He.
The dashed lines correspond to the single nucleon current
calculations. The results including MEC (and in the case
of the charge form factor also the $(p/m)^2$ corrections)
are shown with the solid line.
The data are from \protect\cite{elasticdata}.
      }
    \end{center}
\end{figure}

\begin{figure}[htb]
  \begin{center}
\epsfig{file=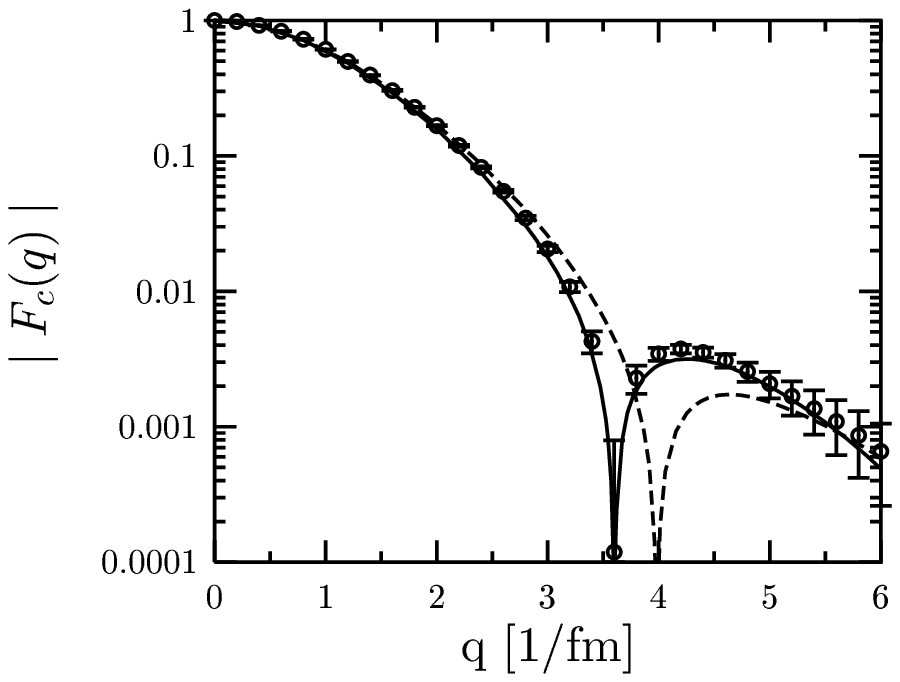,height=4.5cm}
\epsfig{file=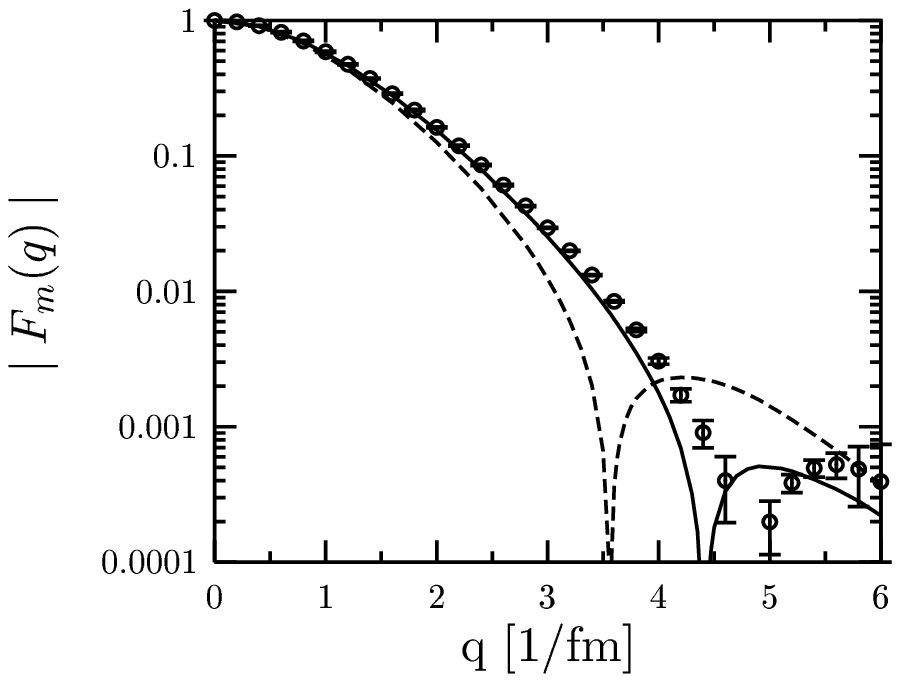,height=4.5cm}
    \caption{\label{fig5}
The same as in Fig.~\protect\ref{fig4} for $^3$H.
      }
    \end{center}
\end{figure}

Figs.~\ref{fig4} and~\ref{fig5}
display the charge
and magnetic form factors in elastic electron scattering on $^3$He and $^3$H, 
respectively.
The agreement with the data for the lower $q$-values is reasonable but
requires two-body densities for the charge form factors and two-body currents 
for the magnetic form factors. Here we are not worried about discrepancies at
the higher $q$-values ($q \ge 3 \ {\rm fm}^{-1}$), where clearly relativistic
effects have to be taken seriously into account.

\subsection{Inclusive Electron Scattering}

\begin{figure}[htb]
  \begin{center}
\epsfig{file=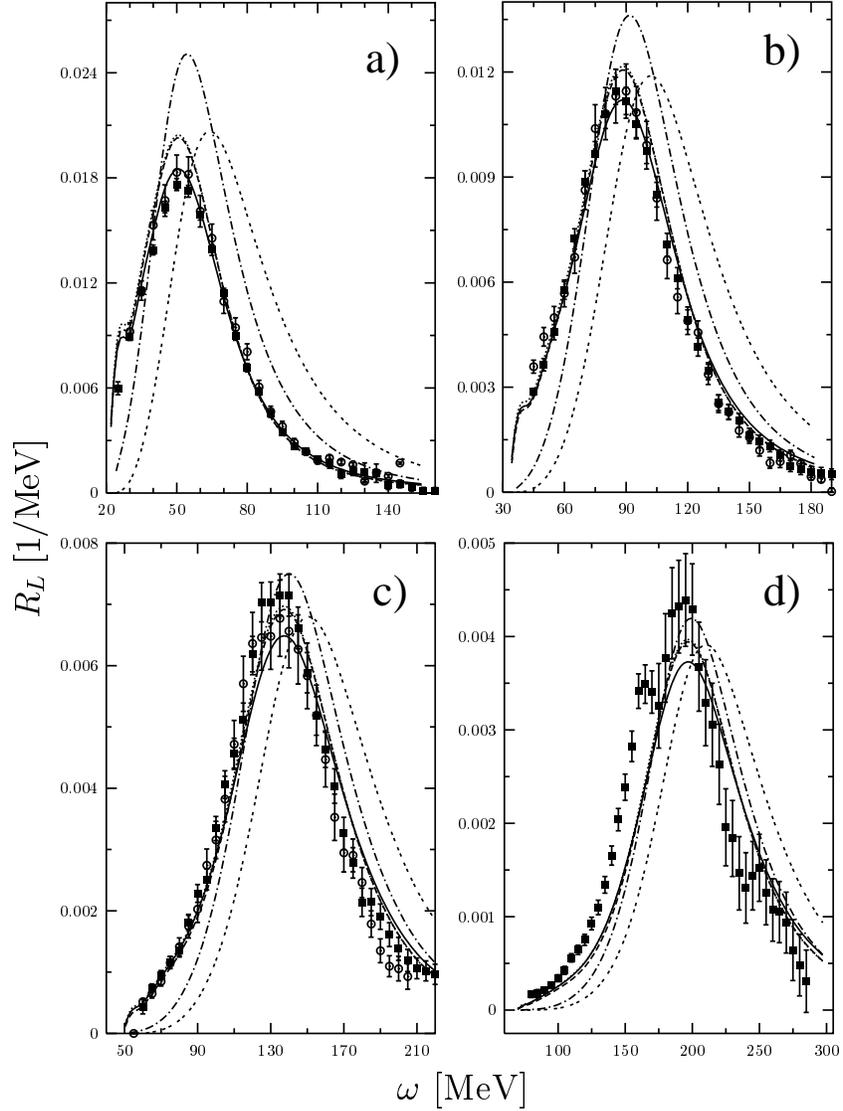,height=15cm}
    \caption{\label{fig6}
The $R_L$ response function against the energy transfer $\omega$
for $Q$= 300 (a), 400 (b), 500 (c), and 600 (d) MeV/c. 
The curves as in Fig.~\protect\ref{fig1} except that the additional
double-dashed line corresponds to the pure (not symmetrized) plane 
wave results.
The data are from \protect\cite{Marchand} (squares) 
and from \protect\cite{Dow} (open circles). The FSI and FSI+MEC curves 
 essentially overlap. 
      }
    \end{center}
\end{figure}

\begin{figure}[htb]
  \begin{center}
\epsfig{file=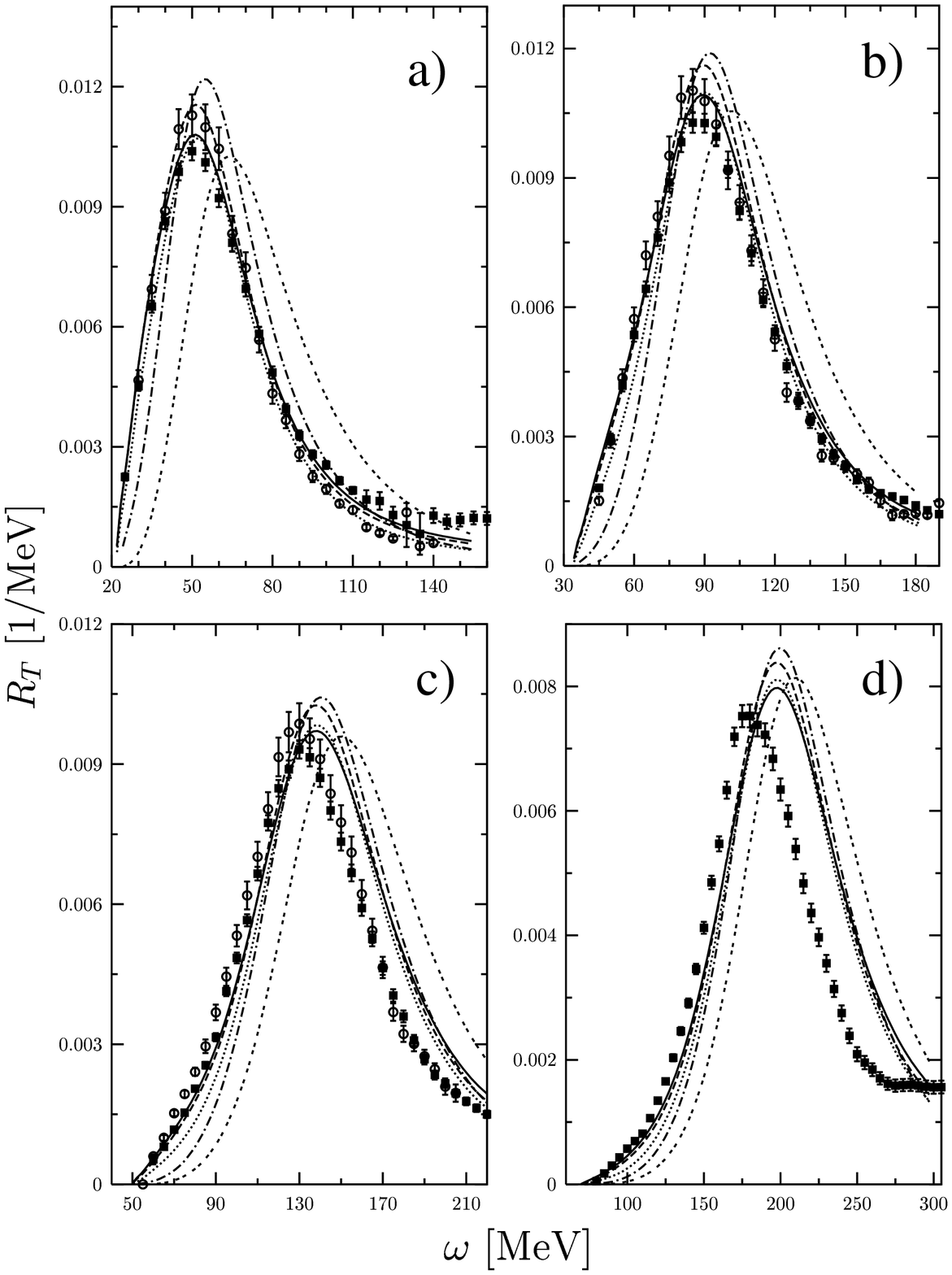,height=15cm}
    \caption{\label{fig6a}
The same as in Fig.~\protect\ref{fig6} for the $R_T$ response function.
      }
    \end{center}
\end{figure}

\begin{figure}[htb]
  \begin{center}
\epsfig{file=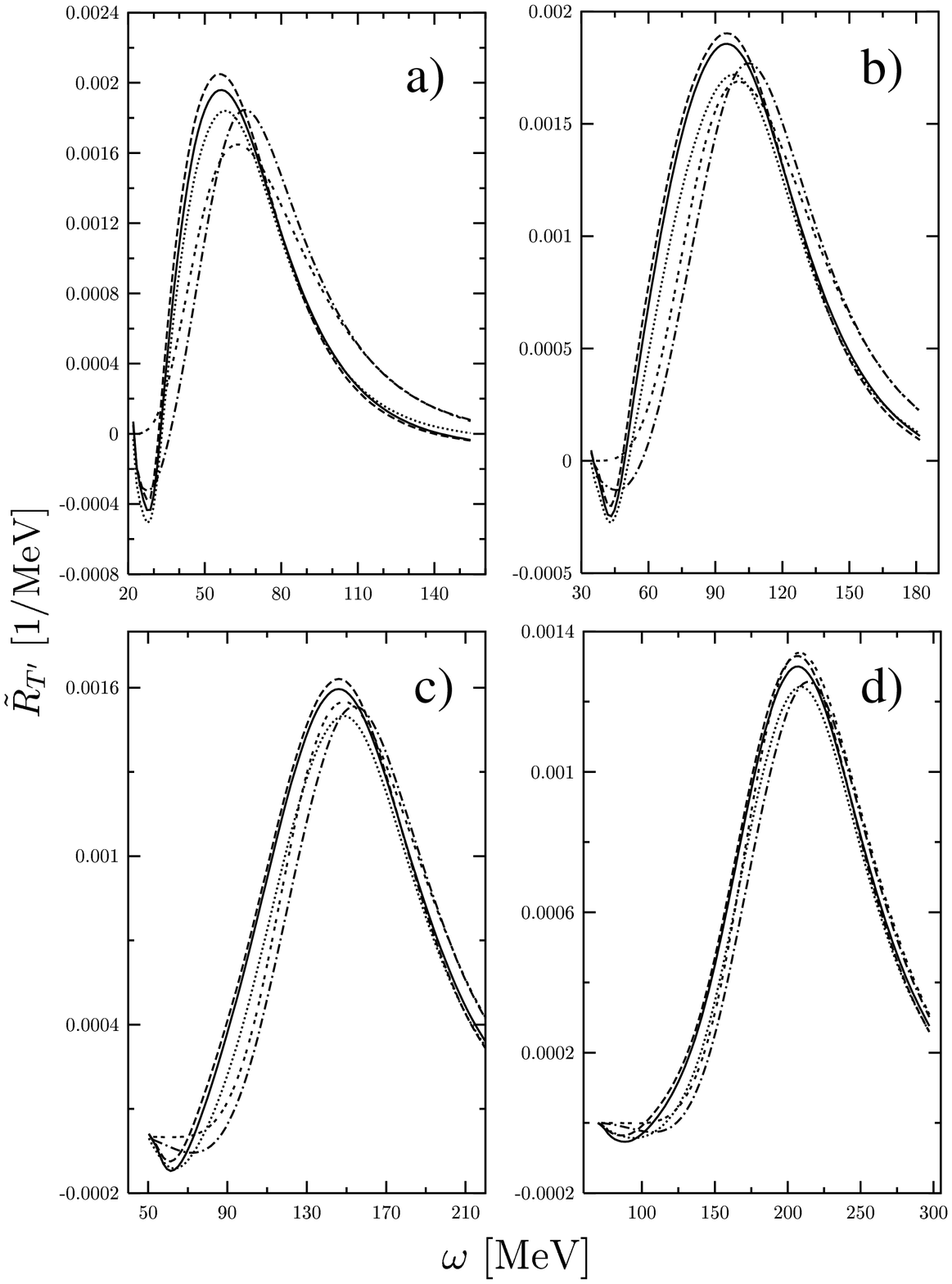,height=15cm}
    \caption{\label{fig6b}
The same as in Fig.~\protect\ref{fig6} for 
the $\tilde{R}_{T'}$ response function.
      }
    \end{center}
\end{figure}

\begin{figure}[htb]
  \begin{center}
\epsfig{file=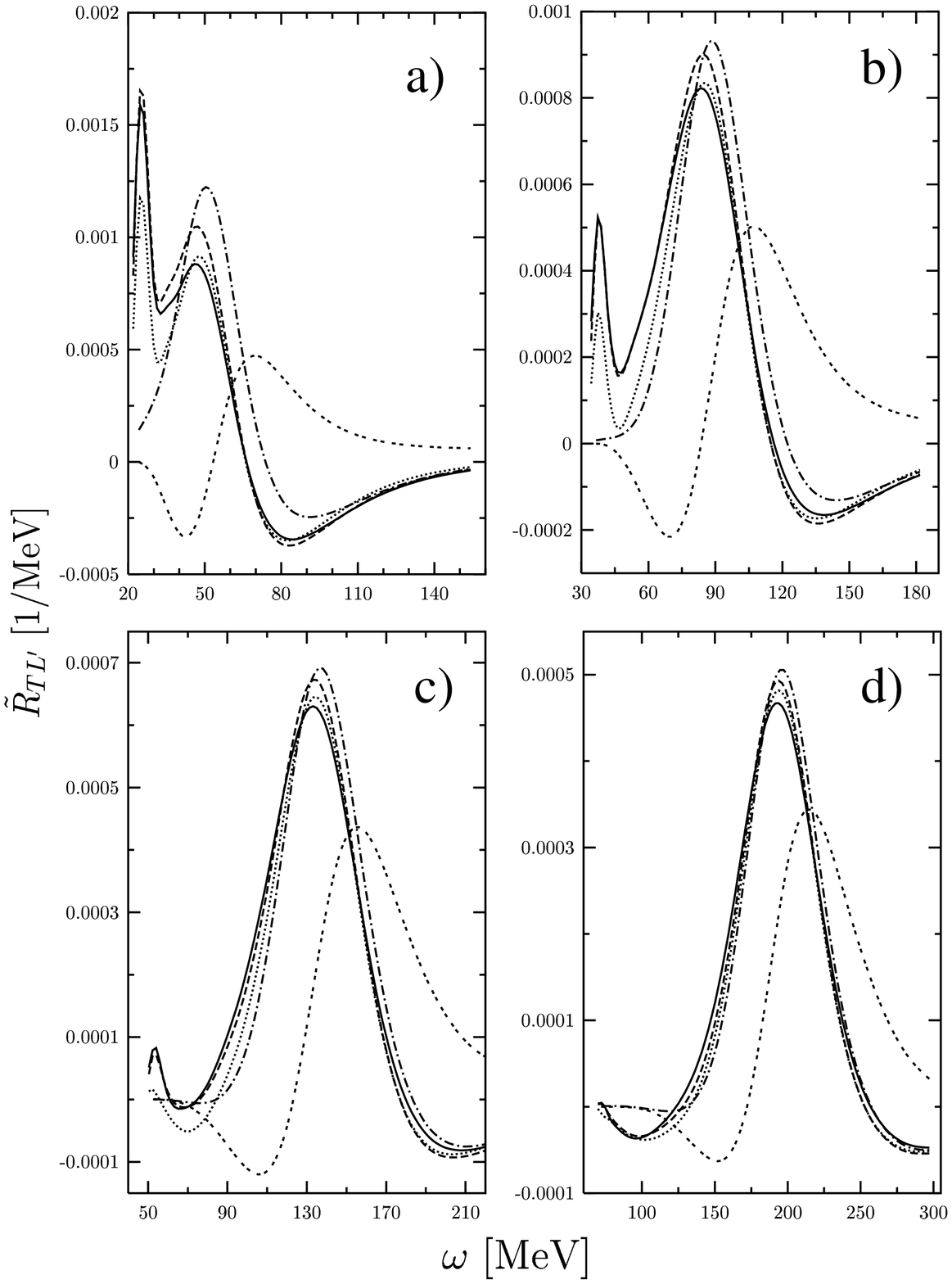,height=15cm}
    \caption{\label{fig6c}
The same as in Fig.~\protect\ref{fig6} 
for the $\tilde{R}_{TL'}$ response function.
      }
    \end{center}
\end{figure}

%

The response functions $R_L$ and $R_T$ are displayed in Figs.~\ref{fig6}
and~\ref{fig6a} 
for $Q$= 300, 400, 500, and 600 MeV/c. In addition to the curves shown 
in  Fig.~\ref{fig1} we included now PWIA.  
This is the  not symmetrized plane wave result, where the absorption
 of the photon takes place on just one nucleon by a single nucleon current
(see Fig.~\ref{fig2}.)
 Later on we also use symmetrized plane wave (PWIAS), where the photon is
 absorbed on all three nucleons by a single nucleon current operator but
all final state interaction is still neglected.
In the approximate treatment of FSI, called FSI23 for short, 
the nuclear matrix element is not
antisymmetrized in the final state (which mathematically is identical to
the assumption that the photon is absorbed only on one of the three nucleons,
not in a symmetrized manner on all three).
FSI stands for the complete calculation with all 
final state interactions 
(exact 3N scattering state $ \langle \Psi_f^{(-)} \mid $). The 
further curves for $R_L$
in Fig.~\ref{fig6} allow for the additional action
of the two-body density operator, and, finally on top of 
that we see the action
of a 3NF. From the figure it is clear that PWIA and FSI23 are rather off.
FSI is extremely important but misses the data. The contribution of the
two-body density is marginal, but the action of the 3NF is quite substantial 
and leads to a nice agreement with the data at $Q$= 300 and 400 MeV/c. At 
the higher $Q$-values relativity is needed. We expect that relativistic 
kinematics will shift theory to the right position. 

In the case of $R_T$
an interesting interplay between MEC's and 3NF effects occurs.
While MEC-effects shift the FSI results upwards in the peak region, 
additional 3NF
effects shift them down again. Unfortunately the two available data sets 
at $Q$= 300  MeV/c are  controversial and do not allow to decide 
about agreement or disagreement between the theory and the data.
In view of that interesting interplay precise new data in the
peak region would be very desirable. On the theory side, 
it would be interesting to investigate the role of
possible 3N currents related to the 3NF used. 
Also here relativistic 
kinematics is required, at least at the two higher $Q$-values. 

The two response functions going with the
helicity of the polarized electron beam 
(occurring in Eq.~(\protect\ref{eq43}))
are equally sensitive to FSI, MEC
and
3NF effects and are displayed in Fig.~\ref{fig6b} and~\ref{fig6c}. 
While for  $\tilde{R}_{T'}$ MEC and 3NF effects
 move the pure FSI result in the same direction the opposite 
is true for $\tilde{R}_{TL'}$.
 In that  response function  the various  effects are quite 
pronounced and one sees a 
strong $\omega$ dependent structure. Therefore data would be 
extremely valuable
 to test all the dynamical ingredients.


Some experimental
information is, however, already available in form of asymmetries
\begin{eqnarray}
A \equiv
{
{
  {{d^3 \sigma } \over {d \Omega_{e'} d E'} } \Bigl \vert _{h=+1}
 -{{d^3 \sigma } \over {d \Omega_{e'} d E'} } \Bigl \vert _{h=-1}
}
\over
{
  {{d^3 \sigma } \over {d \Omega_{e'} d E'} } \Bigl \vert _{h=+1}
 +{{d^3 \sigma } \over {d \Omega_{e'} d E'} } \Bigl \vert _{h=-1}
}
}
\nonumber \\
= -{ {v_{T'} \tilde R_{T'} \cos{\theta ^{*}} + v_{TL'} \tilde R_{TL'}
   2 \sin{\theta ^{*}} \cos{\phi ^{*}}  } \over { v_{L} R_{L} + v_{T} R_{T}}}
\label{eq43}
\end{eqnarray}

\begin{figure}[!ht]
  \begin{center}
\epsfig{file=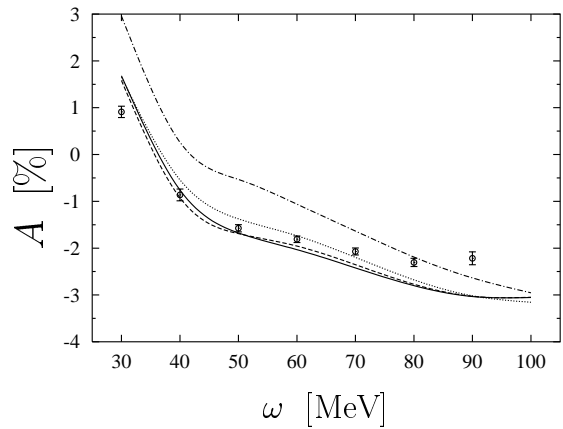,height=6cm}
    \caption{\label{fig8}
            The asymmetry $A$ of Fig.~\protect\ref{fig1}
            together with data from~\protect\cite{Xu.00} 
for $q^2$= 0.1 ${\rm (GeV/c)}^2$. Curves as in Fig.~\protect\ref{fig1}.
      }
    \end{center}
\end{figure}

\begin{figure}[htb]
  \begin{center}
\epsfig{file=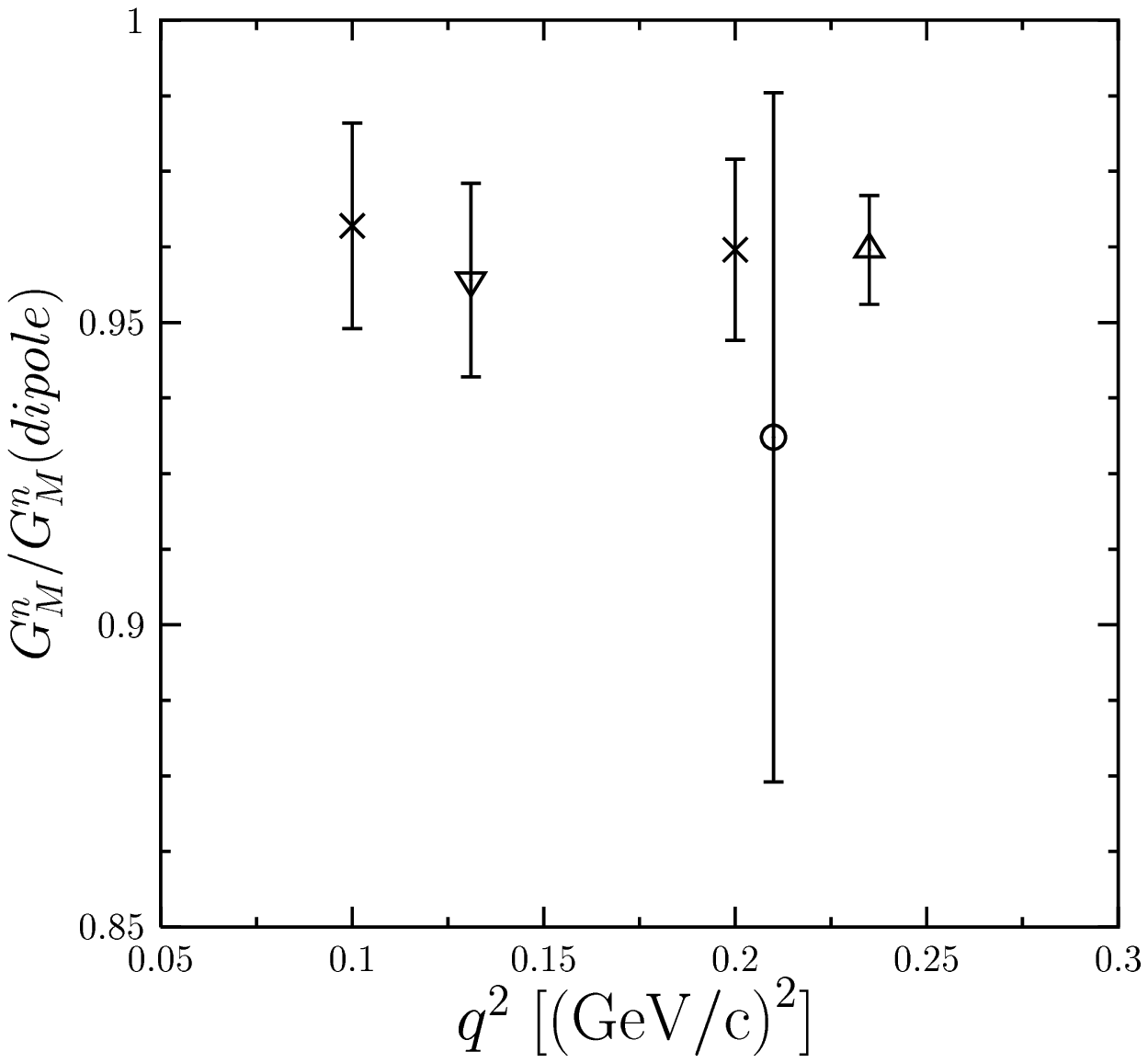,height=6cm}
    \caption{\label{fig7.1}
$G_M^n$-values extracted from different measurements
on the deuteron
(
\protect\cite{Anklin94} ($\bigtriangledown$),
 \protect\cite{Anklin98} ($\bigtriangleup$))
and on $^3$He
(\protect\cite{Gao94} ($\bigcirc$),
 \protect\cite{Xu.00} ($\times$)).
 For the sake of visibility the two deuteron results
 ($\bigtriangledown$ and $\bigtriangleup$) are shifted sidewards
 but belong to $q^2$= 0.1 and 0.2 ${\rm (GeV/c)}^2$, respectively.
      }
    \end{center}
\end{figure}

This well known expression shows the explicit $\theta^*$ and $\phi^*$
 dependence of $A$.
 In Fig.~\ref{fig8} we redisplay the theoretical results 
of Fig.~\ref{fig1} now 
with the data~\cite{Xu.00} taken around $\theta^* = 0^\circ$. 
This allowed us to extract the magnetic neutron form
factor at two $q^2$-values, 0.1 and 0.2 ${\rm (GeV/c)}^2$~\cite{Xu.00}.
The well known property, $\tilde{R}_{T'} \propto (G_M^n)^2$, 
valid in PWIA~\cite{inclusive2}, is lost if the full dynamics is acting, 
nevertheless sufficient sensitivity to $G_M^n$ survives.
The new results for $G_M^n$ extracted from $^3$He  agree
perfectly well with the values extracted from the deuteron~\cite{deuteron} 
(see Fig.~\ref{fig7.1}).

\begin{figure}[!ht]
  \begin{center}
\epsfig{file=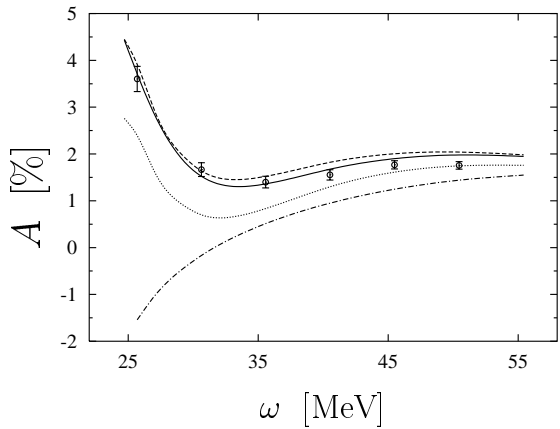,height=6cm}
    \caption{\label{fig9}
            The asymmetry $A$ at $q^2$= 0.1 ${\rm (GeV/c)}^2$
             around $\theta^* = 135^\circ$. 
          The curves as in Fig.~\protect\ref{fig8}
            are shown together with data from~\protect\cite{Xiong}.
      }
    \end{center}
\end{figure}

At the
higher $q^2$-values also measured in~\cite{Xu.00}, our dynamical ingredients
failed, which points to relativistic effects in kinematics and dynamics. In a
following experiment~\cite{Xiong} the formula (\ref{eq43}) has been checked 
around $\theta^* = 135^\circ$, 
where both response functions contribute. The   result is
quite satisfying as shown in one example in Fig.~\ref{fig9}. Nevertheless 
a systematic
experimental study of $\tilde{R}_{T'}$ and $\tilde{R}_{TL'}$ is 
very much needed.

\subsection{The pd-breakup of $^3He$}

The pd breakup process shows two prominent
structures, the proton and deuteron knockout peaks. It is well known that in
PWIA the process
\begin{figure}[!ht]
  \begin{center}
\epsfig{file=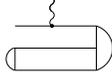,height=1.0cm}
    \caption{\label{fig9.1}
            The diagram corresponding to PWIA.
      }
    \end{center}
\end{figure}
 in parallel kinematics shown in Fig.~\ref{fig9.1} leads to a peak in the
angular distribution of the proton, while the two processes
\begin{figure}[!ht]
  \begin{center}
\epsfig{file=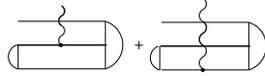,height=1.0cm}
    \caption{\label{fig9.2}
            The diagrams corresponding to the rest of PWIAS.
      }
    \end{center}
\end{figure}
shown in Fig.~\ref{fig9.2} lead to a peak in the angular 
distribution of the deuteron. The three diagrams together correspond 
to taking a fully antisymmetrized final state, which is denoted by PWIAS. The 
diagram in Fig.~\ref{fig9.1} and the two in Fig.~\ref{fig9.2} 
 contribute selectively either to one or to the other peak.
  It is of
interest to investigate rescattering processes for both peaks. This is
displayed in two kinematical examples in Figs.~\ref{fig10} and~\ref{fig11}. 
\begin{figure}[!ht]
  \begin{center}
\epsfig{file=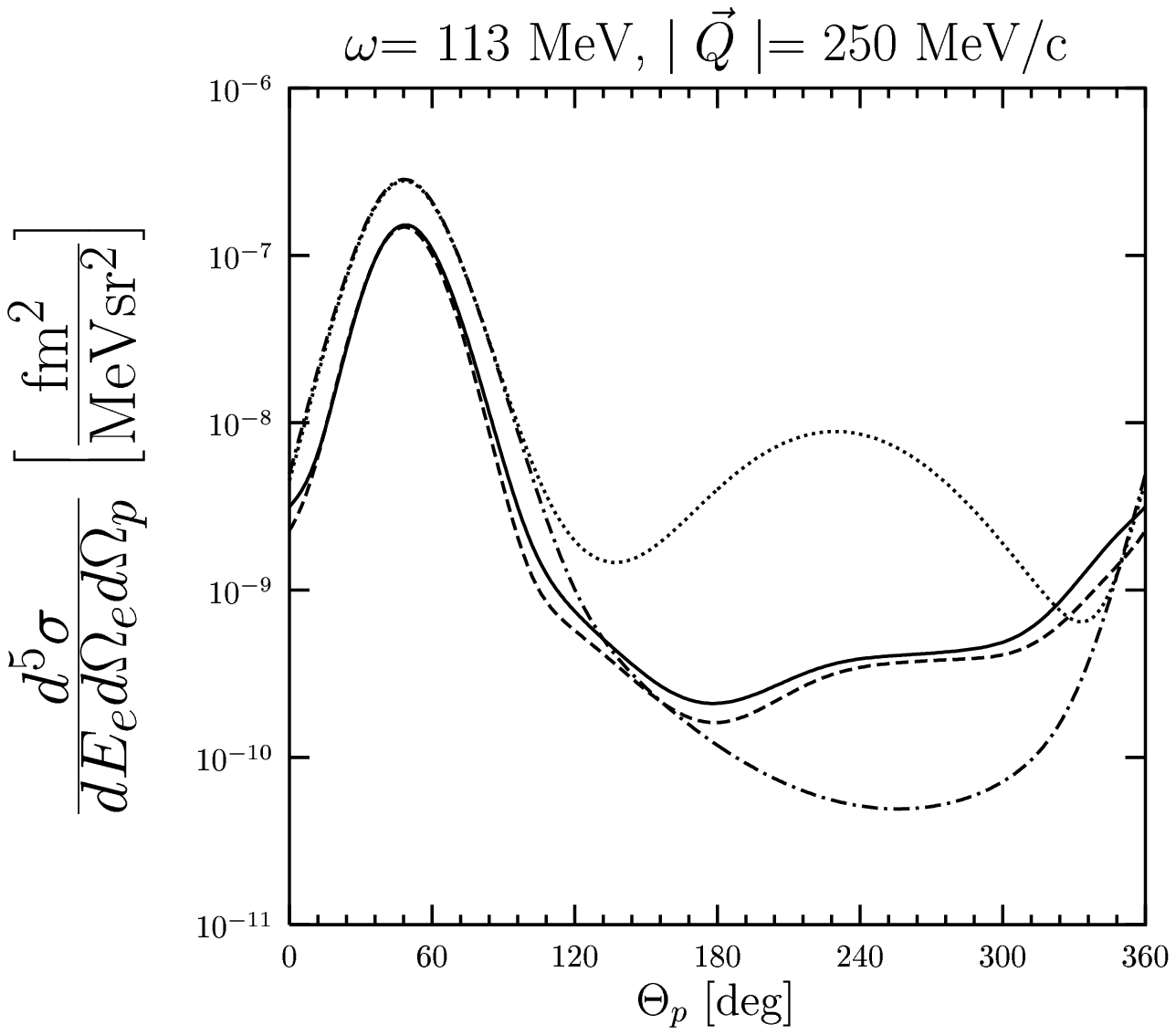,height=5.5cm}
\epsfig{file=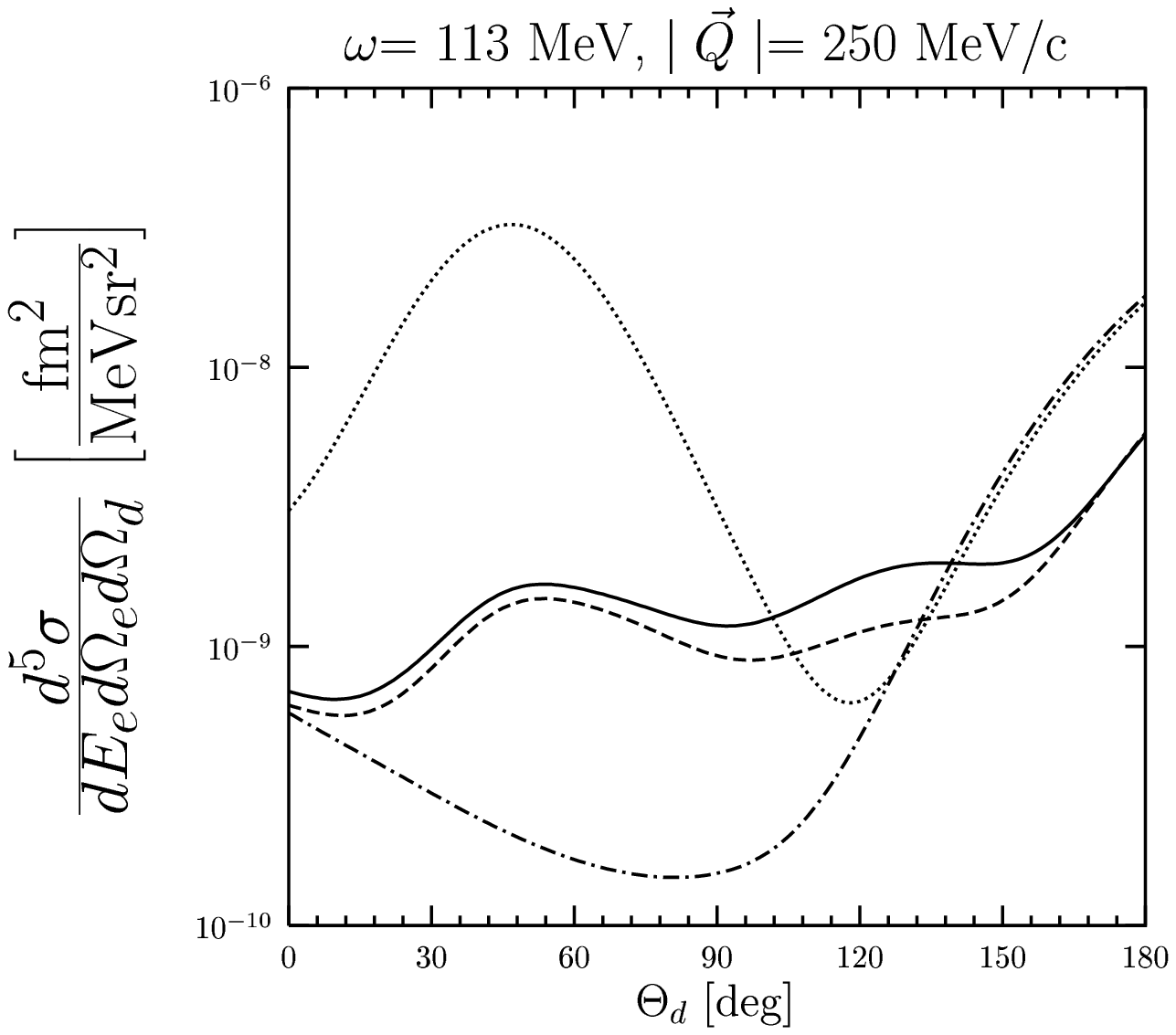,height=5.5cm}
    \caption{\label{fig10}
     The proton (left) and deuteron (right) angular distributions
    for the two-body breakup of $^3$He. 
    The curves depict PWIA (dash-dotted) and PWIAS  (dotted) results, 
            the full calculations without MEC (dashed) and
            the full calculations which incorporate both $\pi$- 
and $\rho$-like
            MEC (solid).
            AV18 is used as the NN potential.
            }
    \end{center}
\end{figure}

\begin{figure}[!ht]
  \begin{center}
\epsfig{file=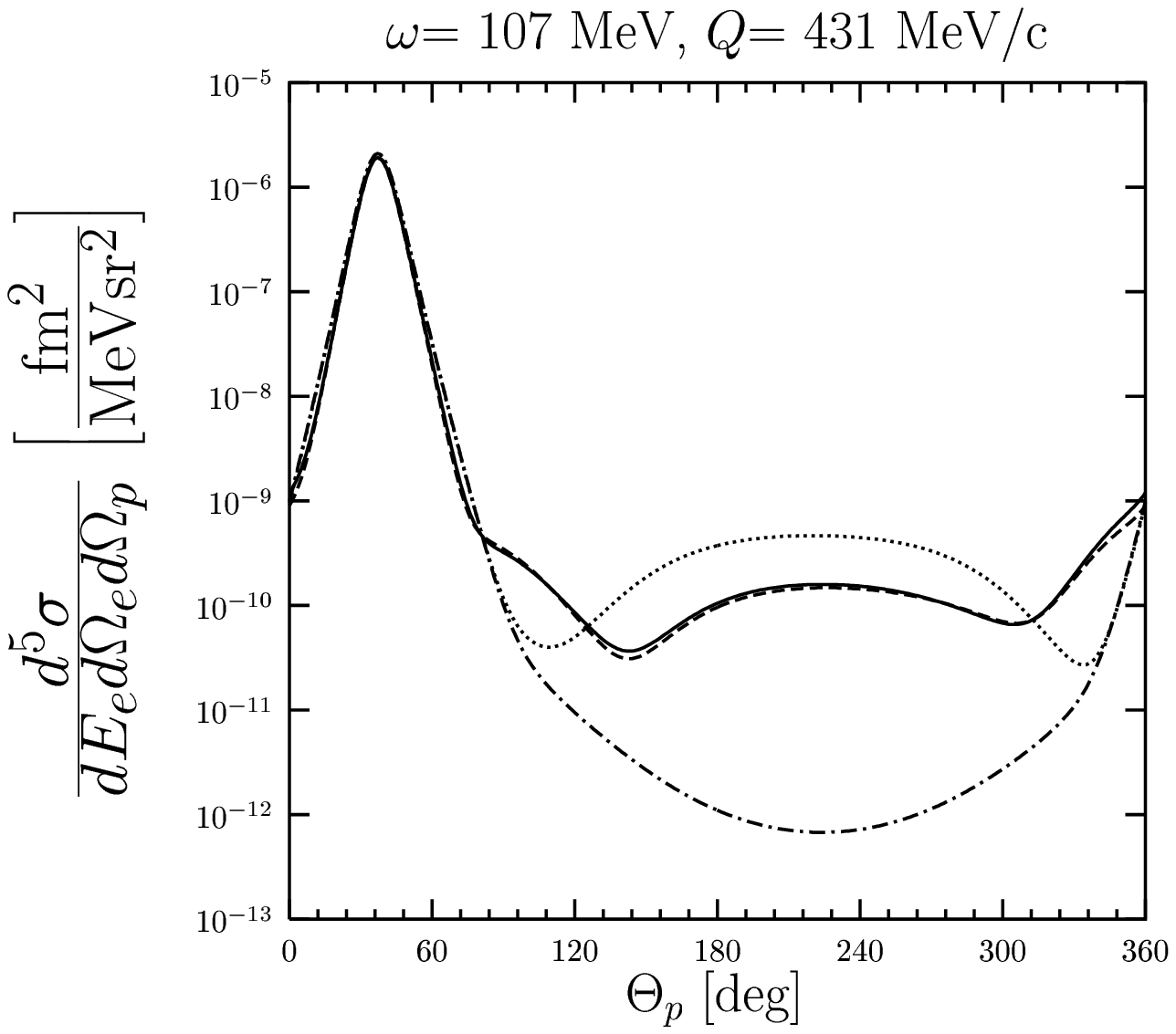,height=5.3cm}
\epsfig{file=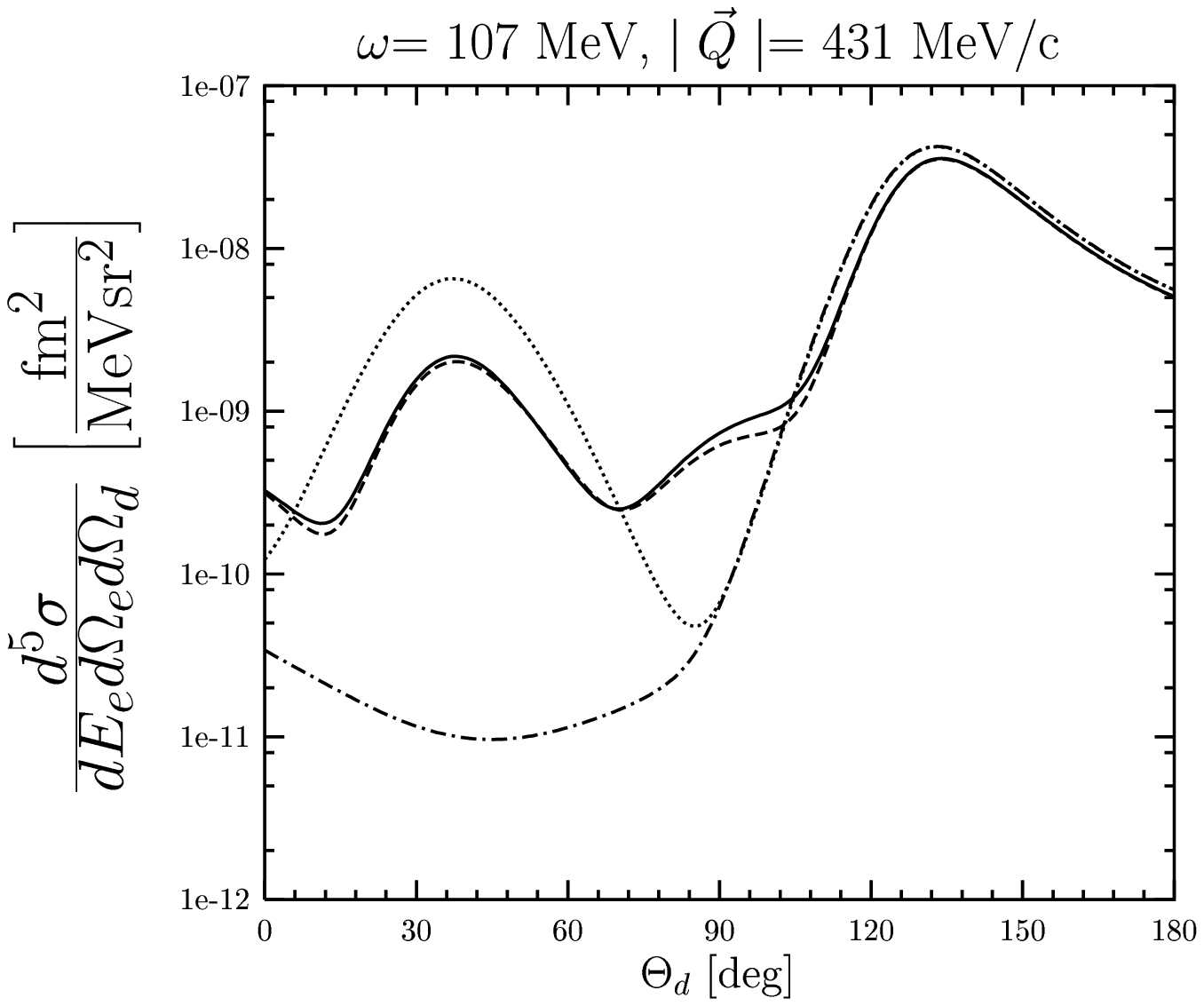,height=5.3cm}
    \caption{\label{fig11}
     The same as in Fig.~\protect\ref{fig10} but for a different 
     ($\omega, Q$) pair.
      }
    \end{center}
\end{figure}

For $Q$= 250 MeV/c we see in PWIA just the p-knockout peak around 
$\theta_p$= 60$^\circ$.
PWIAS shows an additional peak around $\theta_p$= 240$^\circ$ but
its two additional (beyond PWIA) diagrams do not contribute 
around 60$^\circ$.  The deuteron peak 
is clearly visible when plotted against the 
deuteron scattering angle $\theta_d$.  
For the $Q$-value of Fig.~\ref{fig10} FSI effects are strongly visible
in both peaks and also MEC effects to some extent. This is different  
at $Q$=431 MeV/c (see Fig.~\ref{fig11}), where only the deuteron peak
is strongly affected
by FSI whereas the p-peak can be described quite well by the most
simple process, PWIA. For that $Q$-value MEC-effects are nearly negligible.

\begin{figure}[!ht]
  \begin{center}
\epsfig{file=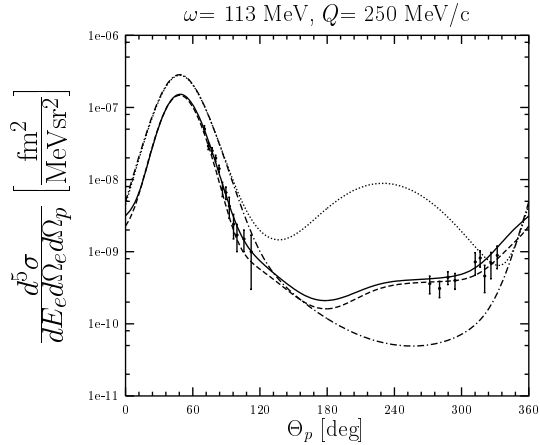,height=6cm}
    \caption{\label{fig12}
     The whole proton angular distribution in the same kinematics as chosen
in Fig.~\protect\ref{fig10}. The data are from [20]. 
The curves as in Fig.~\ref{fig10}.
      }
    \end{center}
\end{figure}

 It would be very desirable to measure both peak regions in one
and the same experiment (the same virtual photon). 
To the best of our knowledge
the two regions have only been investigated in separate 
experiments. We show one 
example in Fig.~\ref{fig12}. 
Even for p-knockout only a part of the peak region has been 
covered by the data. Precise data covering the
whole region would be desirable at low and somewhat higher $Q$- values. At
least at the higher $Q$-values theoretical predictions should 
be correct since the simple diagram in PWIA is just an overlap of the
deuteron and the $^3$He state at relatively low deuteron momenta. At
the low $Q$ - values FSI will play a role but the nuclear forces used describe
pd scattering quite well and MEC effects are very small. 
Therefore also in this case agreement to the
data should be expected. These expectations should be verified by comparison 
to data.

\begin{figure}[!ht]
  \begin{center}
\epsfig{file=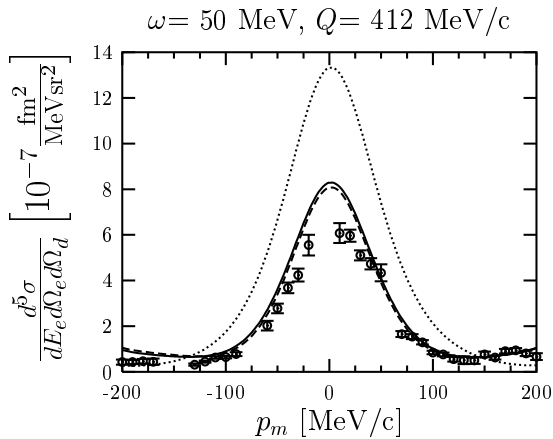,height=6cm}
    \caption{\label{fig13}
     The deuteron knockout peak area as a 
function of the missing (i.e. proton) momentum.
     The data are from \cite{Spaltro}. The curves as in Fig.~\ref{fig10}
      except that the PWIA prediction (totally negligible) is not shown.
      }
    \end{center}
\end{figure}

The situation is even more interesting in the d-peak, 
shown in Fig.~\ref{fig13}, where theory
fails dramatically. Further precise data would be very valuable to
solidify that failure. A confirmation would call for an 
improvement in the currents. Also the role of the 
3N forces in the continuum has to be further studied.

\subsection{The full breakup of $^3He$}


In case of the complete breakup 
$^3{\rm He(e,e'pp)}$ and  $^3{\rm He(e,e'pn)}$ we
refer
to \cite{ppn1} and \cite{ppn2}. Unfortunately the data there were taken in a
kinematical regime where the 3N c.m.energy is above the pion threshold
and serious discrepancies showed up in the comparison to the 
(inadequate)
theory. Data at lower energy transfers would very likely provide
interesting insight into the interplay of NN forces, 3N forces  and MEC's, 
similar to what we found in photon induced break 
up processes \cite{photon1,photon2}.
Here we display two kinematically complete electron induced 
breakup cross sections
with interesting pronounced structures.
\begin{figure}[!ht]
  \begin{center}
\epsfig{file=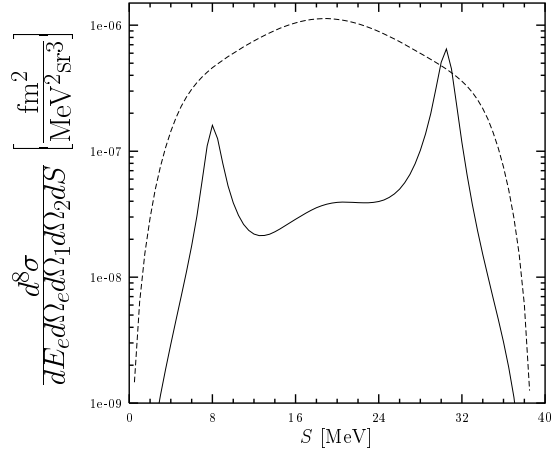,height=6cm}
    \caption{\label{fsipeak}
     The cross section for the three-body breakup of $^3$He for fixed angles 
     of two outgoing nucleons as a function of the arc-length $S$.
     The PWIAS (dashed line) and Full results without MEC and 3NF 
(solid line) are shown.
     The incoming electron energy $E_e$= 390 MeV, 
$\omega$= 37 MeV and $Q$= 100 MeV/c.
     The proton angles are $\theta_1$= 11$^\circ$ and  $\phi_1$= 0$^\circ$, 
     while the second observed nucleon (neutron) is emitted 
     with $\theta_2$= 139$^\circ$ and  $\phi_1$= 180$^\circ$.
      }
    \end{center}
\end{figure}
The strong peaks in Fig.~\ref{fsipeak} arise since
around $S$= 8 MeV and 30 MeV (arc-length along the kinematical locus) 
two nucleons leave with equal momentum vectors. 
For $S \approx $ 8 MeV this happens 
in the neutron-proton (23) subsystem and for 
$S \approx$ 30 MeV in the proton-proton
subsystem (13).  The peaks in Fig.~\ref{planestar}
\begin{figure}[!ht]
  \begin{center}
\epsfig{file=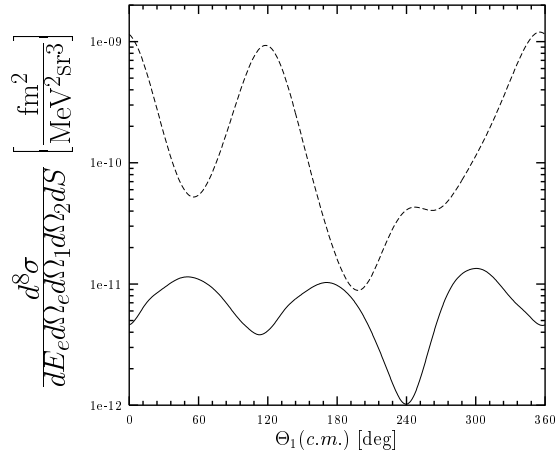,height=6cm}
    \caption{\label{planestar}
     The cross section for the three-body breakup of $^3$He in the so-called 
     ``plane star'' configuration as a function of the c.m. angle of the 
      ejected proton. 
     The PWIAS (dashed line) and Full results without MEC and 3NF (solid line) 
       are shown.
     The incoming electron energy $E_e$ equals 390 MeV, 
$\omega$= 113 MeV and $Q$= 250 MeV/c.
      }
    \end{center}
\end{figure}
shown against the scattering angle
of a nucleon in relation to the virtual 
photon direction correspond to a coplanar
``Mercedes star'' configuration, where the three final nucleons have equal
energies and leave under 120~$^\circ$ pairwise angles.
This is reflected in the repetition of the peaks in steps of 120~$^\circ$.

\subsection{Semi-exclusive nucleon knock-out processes}

Finally, we regard the 
$^3{\rm He(e,e'p)pn}$ and  $^3{\rm He(e,e'n)pp}$
semi-exclusive processes. 
In parallel kinematics and under quasi-free scattering conditions the
spectral function has been an often used tool to analyze the 
data~\cite{more5}. It
is based on the simple picture that the photon is absorbed by the knocked
out nucleon which leaves without any interaction. Only the two spectator
nucleons interact with each other via the NN $t$-operator. This picture
corresponds to the two framed diagrams in Fig.~\ref{fig2}.
In a forthcoming paper \cite{ournewpaper} we shall
investigate in some detail the limitations of that picture 
 in the kinematical region considered in this paper and display
here only some examples. It is easily seen under the simplifying
assumption of the two diagrams that the spectral function is related to
the response functions $R_L$ and $R_T$ as

\begin{eqnarray}
   S(E,k) = \frac12 \, m \, p_{23} \, \frac1{(G_E)^2} 
\int d {\hat p}_{23} R_L (FSI23) \nonumber \\
          = \frac12 \, m \, p_{23} \, \frac{2 m^2}{Q^2 
(G_M)^2}  \int d {\hat p}_{23} R_T (FSI23) .
\end{eqnarray}
 
\begin{figure}[!ht]
  \begin{center}
\epsfig{file=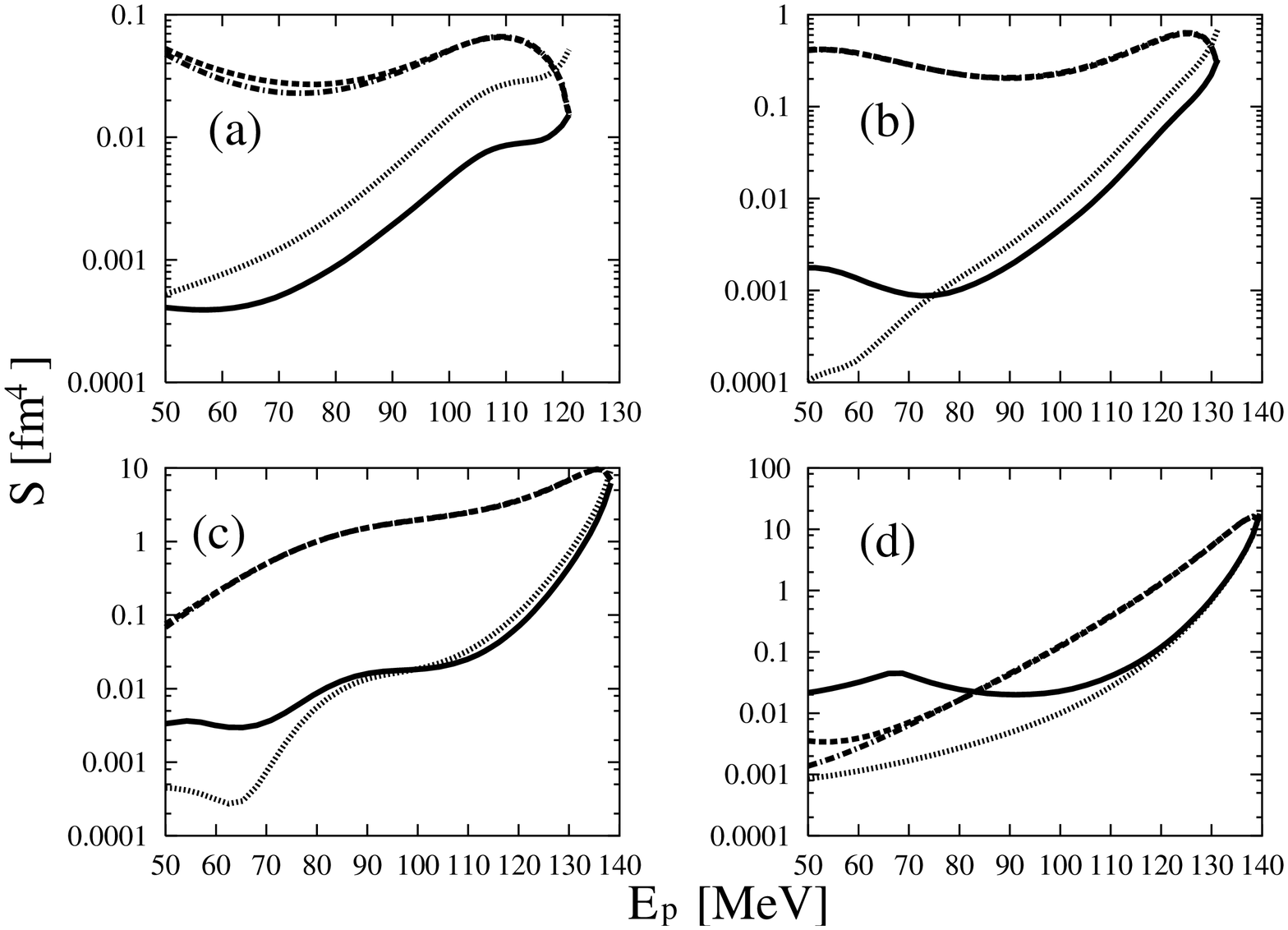,height=9cm}
    \caption{\label{fig.ppnrl}
     The proton spectral function $S(E,k)$ and related $S$-functions 
     from Eq.~(\protect\ref{SSSS})  extracted from $R_L$ 
     for parallel kinematics at $\omega$= 150 MeV and
     four different $Q$ 
values: $Q$= 200 MeV/c (a), $Q$= 300 MeV/c (b), $Q$= 400 MeV/c (c)
     and $Q$= 600 MeV/c (d).
     The PWIA (dash-dotted line), FSI23 (dotted line), PWIAS (dashed line)
     and Full results (solid line) 
are shown as a function of the ejected proton energy.
      }
    \end{center}
\end{figure}

\begin{figure}[!ht]
  \begin{center}
\epsfig{file=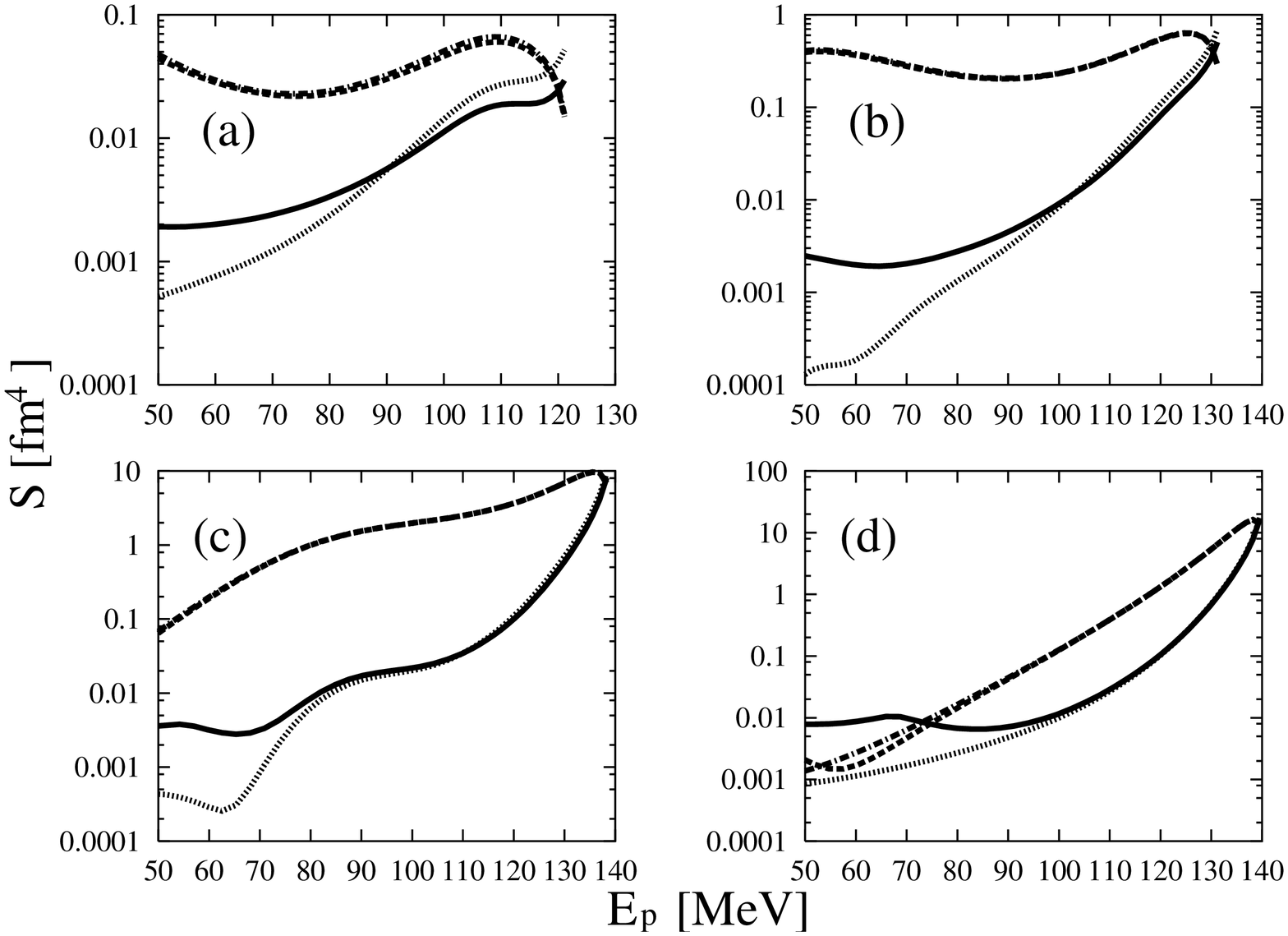,height=9cm}
    \caption{\label{fig.ppnrt}
     The proton spectral function $S(E,k)$ and related $S$-functions
     from Eq.~(\protect\ref{SSSS})  extracted from $R_T$
     for the same conditions as in Fig.~\protect\ref{fig.ppnrl}.
      }
    \end{center}
\end{figure}

\begin{figure}[!ht]
  \begin{center}
\epsfig{file=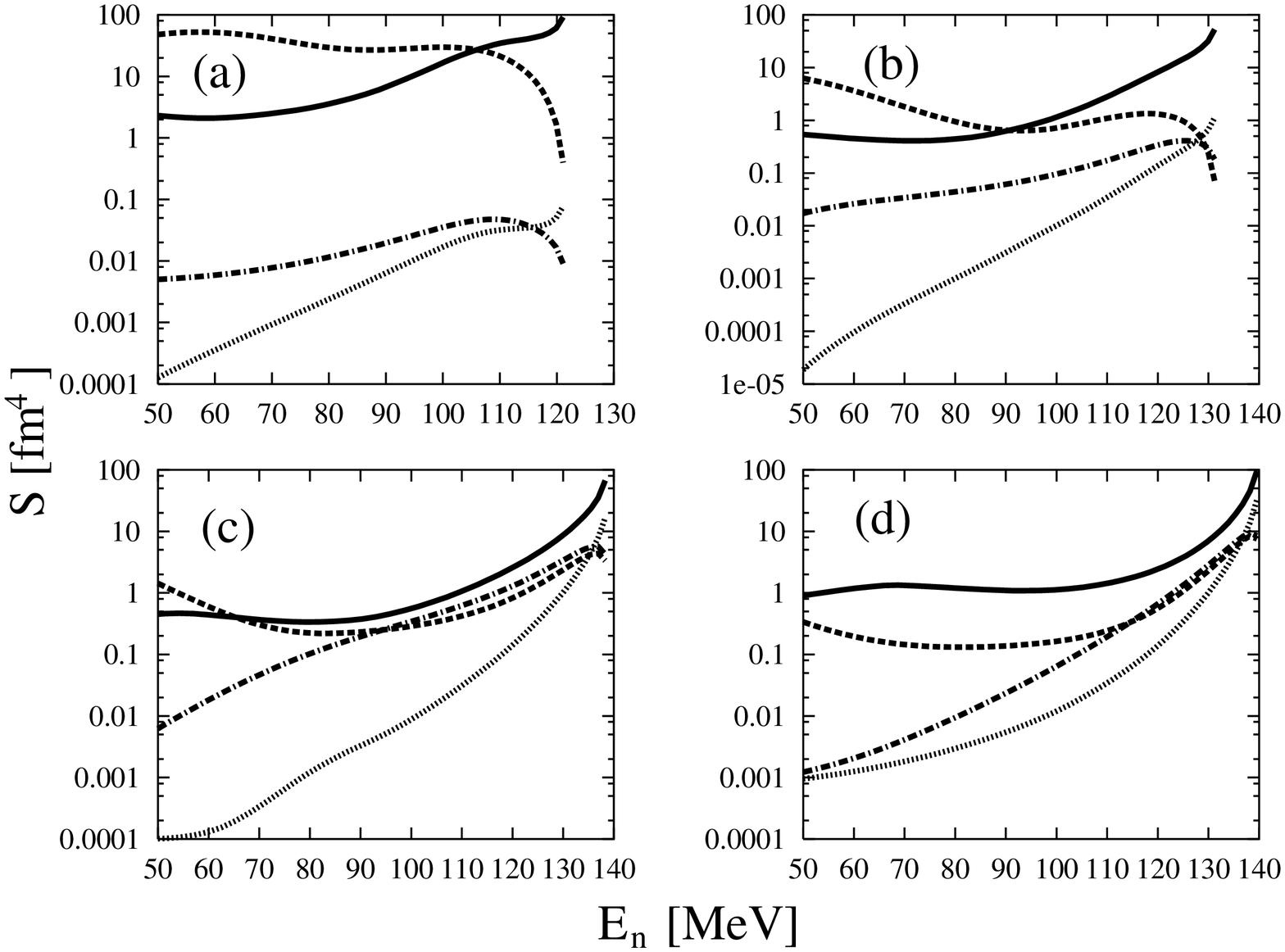,height=9cm}
    \caption{\label{fig.npprl}
     The neutron spectral function $S(E,k)$ and related $S$-functions 
      from Eq.~(\protect\ref{SSSS}) extracted from $R_L$ 
for the same conditions as in Fig.~\protect\ref{fig.ppnrl}.
      }
    \end{center}
\end{figure}

\begin{figure}[!ht]
  \begin{center}
\epsfig{file=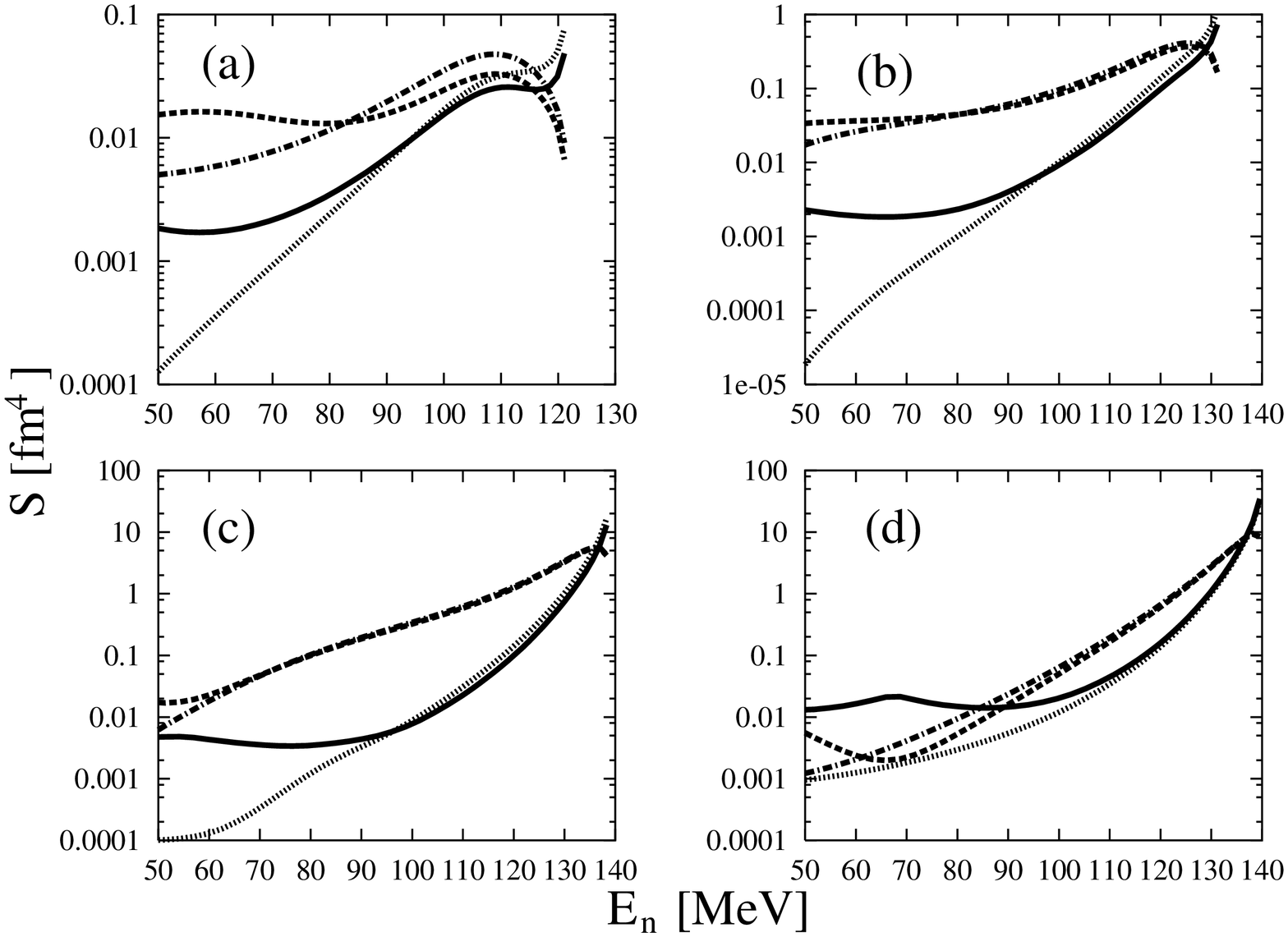,height=9cm}
    \caption{\label{fig.npprt}
     The neutron spectral function $S(E,k)$  and related $S$-functions
      from Eq.~(\protect\ref{SSSS}) extracted from $R_T$
for the same conditions as in Fig.~\protect\ref{fig.ppnrl}.
      }
    \end{center}
\end{figure}

The index FSI23 stands for FSI acting only in the pair 23 when nucleon 1
absorbs the photon. 
 $G_E$ and $G_M$ are the electric and magnetic nucleon form factors and 
$p_{23}$ the relative momentum of the spectator nucleons 2 and~3. 
 These expressions can now be compared to the following expressions 

\begin{eqnarray}
 S(E,k)^{\rm FSI}_{L} = \frac12 \, m \, p_{23} \, \frac1{(G_E)^2} \int d {\hat p}_{23} R_L ({\rm FSI}) 
    \nonumber \\
 S(E,k)^{\rm FSI}_{T} = \frac12 \, m \, p_{23} \, \frac{2 m^2}{Q^2 (G_M)^2}  \int d {\hat p}_{23} R_T ({\rm FSI})
    \nonumber \\
 S(E,k)^{\rm PWIA}_{L} = \frac12 \, m \, p_{23} \, \frac1{(G_E)^2} \int d {\hat p}_{23} R_L ({\rm PWIA}) 
    \nonumber \\
 S(E,k)^{\rm PWIA}_{T} = \frac12 \, m \, p_{23} \, \frac{2 m^2}{Q^2 (G_M)^2}  \int d {\hat p}_{23} R_T ({\rm PWIA})
    \nonumber \\
 S(E,k)^{\rm PWIAS}_{L} = \frac12 \, m \, p_{23} \, \frac1{(G_E)^2} \int d {\hat p}_{23} R_L ({\rm PWIAS}) 
    \nonumber \\
 S(E,k)^{\rm PWIAS}_{T} = \frac12 \, m \, p_{23} \, \frac{2 m^2}{Q^2 (G_M)^2}  \int d {\hat p}_{23} R_T ({\rm PWIAS}) ,
\label{SSSS}
\end{eqnarray}
where $R_{L,T} ({\rm FSI})$, $R_{L,T} 
({\rm PWIA})$ and $R_{L,T} ({\rm PWIAS})$ are the response functions
 calculated with the complete FSI, just using PWIA or 
finally the symmetrized version PWIAS.
For various fixed  $(\omega,Q)$ pairs we show in 
Figs.~\ref{fig.ppnrl}--\ref{fig.npprt}
the quantities $S(E,k)$ as a function of $E_1$, 
the energy of the knocked out nucleon.
To each $E_1$ corresponds uniquely the value of the missing momentum $k$
($\vec k = {\vec p}_1 - {\vec Q}$) and excitation energy 
$ E = p_{23}^2/m$. 
 We display in Table~\ref{tab1}
\begin{table}
\begin{tabular}{ccc} 
 \hline
 $E_1$ (MeV) & $E$ (MeV) & $k$ (${\rm fm}^{-1}$)  \\
 \hline
 &  $Q$=200 MeV/c  & \\
      50.0  &  89.27  &   0.54 \\
      70.0  &  65.24  &   0.82 \\
      90.0  &  40.41  &   1.07 \\
     110.0  &  15.04  &   1.29 \\
     121.7  &   0.00  &   1.41 \\
 \hline
 &  $Q$=300 MeV/c   & \\
      60.0  &  81.94  &   0.18  \\
      80.0  &  60.24  &   0.44  \\
     100.0  &  37.55  &   0.68  \\
     120.0  &  14.15  &   0.89  \\
     131.9  &   0.00  &   1.00  \\
 \hline 
 &  $Q$=400 MeV/c  & \\
      50.0  &  89.95  &   0.47   \\
      60.0  &  81.18  &   0.33   \\
      80.0  &  62.24  &   0.06   \\
     100.0  &  41.99  &   0.17    \\
     120.0  &  20.80  &   0.38   \\
     139.0  &   0.00  &   0.56   \\
 \hline 
&   $Q$=600 MeV/c  & \\
      50.0  &  69.34  &   1.49  \\
      70.0  &  57.28  &   1.20  \\
      90.0  &  42.78  &   0.96  \\
     110.0  &  26.65  &   0.74  \\
     130.0  &   9.30  &   0.54  \\
     140.3  &   0.00  &   0.44  \\
\end{tabular}
\caption[]{The relation between $E_1$, the energy of the nucleon
ejected parallel to $\vec Q$, and \mbox{($E$, $k$)},
the arguments of the spectral function $S(E,k)$
for four different values of $Q$.
In all cases the energy transfer $\omega$= 150 MeV.
}
\label{tab1}
\end{table}
the kinematical variables underlying Figs.~\ref{fig.ppnrl}--\ref{fig.npprt}.
This together with $S(E,k)$ given in Figs.~\ref{fig.ppnrl}--\ref{fig.npprt}
is a substitute for plotting $S(E,k)$
over the $(E,k)$ plane.

 Let us first regard the p-knockout in Figs.~\ref{fig.ppnrl} 
and~\ref{fig.ppnrt}. For $R_L$ only in case
of $Q$= 600 MeV/c  the use of the spectral function $S$ 
is a good approximation at the upper end of $E_1$,
where the FSI (solid line) and FSI23 (dotted line) predictions coincide. 
For that restricted energy range in $E_1$ the absorption of the photon 
by nucleons 2 and 3, as added in for PWIAS, provide a negligible 
contribution to the PWIA result. This is of course different at smaller 
$E_1$ values (not shown). 
 The approximate use of $S$ in case of $R_T$ shown in Fig.~\ref{fig.ppnrt}  
is more justified. In 
all cases PWIA 
or PWIAS would be meaningless.

In case of n-knockout shown in Figs.~\ref{fig.npprl} and~\ref{fig.npprt} 
FSI is much more present for $R_L$ than for p-knockout. 
Here the use of the spectral function would be quite erroneous. In case of 
$R_T$, however, the situation is similar as for p-knockout.   
  Now the results
for PWIA and PWIAS differ which is due to the strong absorption on the proton.

Systematic measurements related to those examples would be very helpful to
test the dynamics. 
This would be especially gratifying in case of the neutron knock out which should be well
understood in order to extract the electric form factor of the neutron in
a reliable manner. We refer to \cite{een} where the process 
$\overrightarrow{^3{\rm He}}(\vec{e},e'n)$
has been studied extensively with the result that for the kinematical
region considered here FSI is extremely important.
Outside the $(\omega-Q)$ domain considered here the use of $S(E,k)$
might be more favorable~\cite{more5} but its 
justification requires a relativistic treatment.

\section{Outlook}
\label{sec:4}
 We showed for various observables in electron induced inelastic processes
 on $^3$He that they challenge our present day understanding of nuclear forces
 and currents. The 3N system is an ideal laboratory  since all reactions can
 be safely calculated in the Faddeev scheme and no uncontrolled approximations
 blur the comparison between theory and experiment. There are quite a few
 discrepancies using the modern NN forces and the still too restricted set
 of 3N forces and exchange currents. 
For more examples we refer to 
\cite{Xu.00,Spaltro,ppn2,more1,more2,more3,more4}.
Dedicated and precise experiments, some of
 which are pointed out, would certainly help to lay a solid ground of data,
 which present and future theory have to describe.

 One can expect that the theoretical approach via effective field theory
 constrained by chiral symmetry will lay a corresponding sound basis in theory,
 since NN and 3N forces as well as exchange currents follow from one and the
 same underlying Lagrangian and are therefore consistently defined.
 This upcoming theoretical formulation will call even more for adequate data.

\vspace{2cm}

{\centerline{ACKNOWLEDGMENTS}

\vspace{0.51cm}

\noindent
This work was supported by the Polish Committee for Scientific Research 
under grant no. 2P03B00825, by the NATO grant no. PST.CLG.978943
and by DOE under grants nos. DE-FG03-00ER41132 and DE-FC02-01ER41187.
The numerical calculations have been performed on the cray SV1 and T3E
of the NIC in J\"ulich, Germany. 
R.S. thanks the Foundation for Polish Science for financial support.

\end{document}